\documentclass[aps,showpacs,preprint,superscriptaddress,nofootinbib]{revtex4}
\font\rsfs=rsfs10.tfm scaled 1000
\usepackage{amsmath,amscd,amssymb,euscript,eufrak}
\usepackage{graphicx}
\usepackage{graphics}
\usepackage{epsfig}
\usepackage{ifthen}
\usepackage{verbatim}

\DeclareMathAlphabet{\mathpzc}{OT1}{pzc}{m}{it}
\def\dia{\diamond}
\def\ups{\upsilon}
\def\c #1{\mbox{\rsfs  #1}} 
\def\tsf#1{{\mathsf{ #1}}}

\def\Re{\mathbb{R}}

\def\Gam{\Gamma}          \def\gam{\gamma}
\def\vp{\varphi}          
\def\eps{\epsilon}        \def\lam{\lambda}
         \def\om{\omega}        \def\Om{\Omega}
\def\eps{\epsilon}        \def\tl{\tilde}
\def\a{\alpha}            \def\bt{\beta}
\def\Del{\Delta}          \def\del{\delta}
\def\kap{\kappa}
\def\co{{{}^\ast}}        \def\cov{{{}_\ast}}
\def\Sig{\Sigma}          \def\we{\wedge}
\def\sig{\sigma}
\def\ra{\rangle}          \def\la{\langle}      
\def\ord#1{{}^{(#1)}}
\def\hph#1{{\hphantom{#1}}}      
\def\nonu{\nonumber}
\def\beq{\begin{equation}}
\def\eeq{\end{equation}}
\def\beg#1#2{ \begin{#1}   #2 \end{#1}}
\def\beqm#1{ \beg{equation} {#1} }
\def\begg#1{ \begin{gather} #1 \end{gather} }
\def\ndef#1{\begin{defin}$\!\!\!$.#1 \end{defin}}

\newtheorem{theo}{{\em Theorem}}
\newtheorem{defin}{{\em Definition}}

\def\SUM#1#2{\mbox{$\sum\limits_{#1}^{#2} $} }
\def\Re{\mathbb{R}}
\def\pr{^\prime}  
\def\llist#1{\begin{list}{}{ \leftmargin=10mm \labelsep=2mm  \labelwidth 4mm
\parsep=1mm  \topsep =1mm \itemsep =0mm} #1  \end{list} }
\def\itt#1{ \item[$(#1)$]  }
\begin{document}
\def\ss #1{ \hskip -1.5pt #1 \hskip -1.5pt }
\def\pl{\partial}
\title
{Geometrical crossover in two-body systems in a magnetic field}
\author{M. Cerkaski}
\affiliation{Institute of Nuclear Physics PAN, Department of Theory
of Structure of Matter, 31-342 Cracow, Poland}
\affiliation{Bogoliubov Laboratory of Theoretical Physics,
Joint Institute for Nuclear Research, 141980 Dubna, Russia}
\author{R.\ G.\ Nazmitdinov}
\affiliation{Bogoliubov Laboratory of Theoretical Physics,
Joint Institute for Nuclear Research, 141980 Dubna, Russia}
\affiliation{Departament de F{\'\i}sica,
Universitat de les Illes Balears, E-07122 Palma de Mallorca, Spain}
\date{\today}
\begin{abstract}
An algebraic approach is formulated in the harmonic approximation
to describe a dynamics of two-fermion systems,
confined in three-dimensional axially symmetric parabolic potential, in an external
magnetic field.
The fermion interaction is considered in the form
$\c U_{M}( r )= \alpha_{ M}\,r^{-M} $  $(\alpha_{M}>0, M>0)$.
The formalism of a semisimple Lie group is applied to analyse symmetries
of the considered system. Explicit algebraic expressions are derived
in terms of system's parameters and the magnetic field strength
to trace the evolution of the equilibrium shape.
It is predicted that the interplay of classical and quantum correlations may lead to
a quantum shape transition from a lateral to a vertical localization of fermions
in the confined system.
The analytical results demonstrate a good agreement with
numerical results for two-electron quantum dots in the magnetic field,
when classical correlations dominate in the dynamics.
\end{abstract}
\maketitle
\section{Introduction}
Symmetry breaking phenomena play important role in the interpretation
of various physical properties of finite many-body systems, for example,
such as nuclei \cite{cej} and metallic grains \cite{kres}.
There exists a type of symmetry transformation that is specific for finite
systems. This is a shape symmetry breaking, when a finite system, under
varying external or internal parameters, exhibits the change of its shape.
This change can be spontaneous, in the sense that the shape form is not
imposed from outside, but the system acquires the chosen form because it
is energetically profitable. Evidently, due to finite number of particles
quantum fluctuations play essential role in the evolution of
various properties of a system.

Recent progress in nanotechnology opens a broad avenue to study
the interplay between microscopic (quantum) and macroscopic (classical) scales
in mesoscopic systems.
If in a mesoscopic system several particles are
confined by a one-body field, the dynamics of the one-body
field governs the individual motion of the particles.
A natural question is how this is changed if the particles are
influenced by a two-body interaction in addition to the one-body
field.
How, for example, in a mesoscopic system with a few particles moving in
one-body potential could be exhibited a symmetry breaking phenomenon
due to two-body interaction,
related to a shape transition driven  by quantum fluctuations ?
The answer on this question may shed light on the connection between
a shape transition and  a quantum phase transition \cite{Sad} in
finite many-body quantum systems, in general.

To study the combined role of one-body
and two-body interactions on symmetry breaking phenomena,
we shall concentrate on the simplest nontrivial case,
namely, the interacting two-body system. Specifically, we focus on two
identical charged particles (electrons) in a three-dimensional deformed harmonic
oscillator potential under a perpendicular magnetic field
(see, for example, \cite{kai,taut,din,ent}).
It is noteworthy on the fact that semiconductor technology made possible
to fabricate and probe such confined system at different values of the magnetic
field \cite{as,kou}. Consequently, it has stimulated numerous theoretical
studies on two-electron quantum dots (QDs), so-called "artificial He atoms" (see for a recent
review \cite{RM,naz,PR}). For example, a circular dot at arbitrary values of
the magnetic field was studied in various approaches in order to find a closed-form solution
\cite{loz,kand1,grokr}.
Being a simplest nontrivial system, QD He poses a significant
challenge to theorists. Indeed,
using a two-dimensional He QD model, one is able to reproduce a general trend for
the first singlet-triplet (ST) transitions observed in two-electron QDs
under a perpendicular magnetic field.
However, the experimental positions of the ST transition
points are systematically higher \cite{kou,nishi}.
The ignorance of the third dimension is the most evident source
of the disagreement, especially, in vertical QDs \cite{din,ron,bruce,nen3}.

The purpose  of the present paper is to analyse correlation
effects produced by a two-body interaction
in   most general form
$\c U_{M}( r )= \alpha_{ M}\,r^{-M} $  $(\alpha_{M}>0, M>0)$  on the
evolution of the ground and excited states of two-fermion (two-electron)
systems under a perpendicular magnetic field.
Although accurate numerical results for such potentials can be obtained
readily, analytical results are still sought even in this case, because
they provide the physical insight into numerical calculations.
Moreover, analytical results could
establish a theoretical framework for accurate analysis of
confined many-electron systems, where the exact treatment of
a three-dimensional (3D) case becomes
computationally intractable.

The content of the paper is following.
In Sec.II we formulate a general two-body problem
confined by a one-body potential. Specifically, we consider two identical
charged particles (electrons) in the field of the three-dimensional
harmonic oscillator, interacting via a  two-body interaction
$\c U_{M}( r )= \alpha_{ M}\,r^{-M} $  $(\alpha_{M}>0, M>0)$.
We show how the two-body Hamiltonian can be scaled, thus reducing the number
of independent variables in the problem.
This scale is related to an additional symmetry of
the Hamiltonian function  ($G_4$ group).
Sec.III is devoted to the development of the algebraic approach to study
shape transitions induced by the classical component of the total
energy of two-electron system in a magnetic field.
The role of quantum fluctuations is discussed in Sec.IV.
The comparison of the analytical results with numerical calculations \cite{nen3}
is presented Sec.V. Sec.VI summarizes briefly the main results.
In Appendix some technical details of Sec.III are discussed.
\eject
\section{Model}
We consider  the Hamiltonian
\begg{
\label{ham}  H =   \sum_{j=1}^2 \bigg[ \frac{1}{2m^*\!}\, \Big({\bf p}_j -e\,\tsf{ A}_j
  \Big)^{\! 2} + U(\tsf{ r}_j) \bigg] +  \c{U}_M(\tsf{ r}) + H_{\rm spin}.  }
For the perpendicular magnetic field we choose the vector
potential with gauge $ \tsf{ A} = \frac{1}{2}\, \tsf{B} \times \tsf{r} = \frac{1}{2}\,B(-y, x,0)$.
The confining potential is approximated by a 3D axially-symmetric harmonic oscillator
$U(\tsf{r}) = m^*\, [\omega_\rho^2\,(x^2 \!+ y^2) + \omega_z{^2}\,z^2]/2$,
where $\hbar\,\omega_z$ and $\hbar\,\omega_\rho$ are the energy scales of confinement
in the $z$-direction and in the $xy$-plane, respectively.
The term   $H_{\rm spin} \ss = \tfrac 1 2\,  g^*\,\mu_B\,( \vec \sig _1 +
\vec \sig _2)  \cdot  \tsf{B} \ss= g^*\,\mu_B\,\hat \Sig_S\,B\,
(\hat \Sig_{S}= \hat S_z/\hbar)$ describes the Zeeman interaction,
where $\mu_B=e\,\hbar/(2\,m_e) $ is the Bohr magneton (the  SI  system of units).
The interaction between two electrons is chosen in most general form
$\c U_M(r)= \a_M\,r^{-M}$ ($\a_M>0$, $M>0$).
In particular, the Coulomb repulsion between two electrons is
$\c U_{M=1}(r)= \a_{1}/r$ with  $\a_{1} = e^2\,(4\,\pi\,\epsilon_0\,\epsilon_r)^{-1}\ss= \hbar\,c\,\a/\eps_r$,
where $\a=e^2/(\hbar\, c) \approx 1/137.036$ ($4\pi\epsilon_0=1)$
is a constant of the subtle structure, and $\eps_r$ is a relative permittivity.
As an example, we will use the effective mass $m^*=0.067m_e$, the relative
dielectric constant $\varepsilon_r=12$ and the effective Land\'e factor $|g^*|=0.3$ (bulk GaAs values).

For our analysis it is convenient to employ  a transformation of single-particle
canonical variables: $(\tsf{m}_1,\tsf{m}_2)\ss=\tsf{m}_{12} \in M_{12}$
where $\tsf{m}_i\ss=(\tsf{r}_i,\tsf{p}_i)\in \Re^3\otimes \Re^3$, -- to scaled
dimensionless  canonical variables of relative $(\tsf{r},\tsf{p})$ and
center-of-mass $(\tsf{r}\cov, \tsf{p}\cov)$ motions, respectively. In other words,
$\tsf{m}_{12}  \to  ((\tsf{r}\cov,\tsf{p}\cov),(\tsf{r},\tsf{p}))\ss=
(\tsf{m}\cov,\tsf{m}) \ss=\tsf{m}_\dia \in M_\dia$
where  each element  of set  $\{\tsf{p}\cov,\tsf{r}\cov,\tsf{p},\tsf{r}\}$ belongs to $\Re^3$,
and the linear transformation is:
\begg{ \label{kappa} \kappa(L_\dia,\eta)\!: \quad  \binom{ \tsf{r}\cov ~
\tsf{r} }{ \tsf{p}\cov ~ \tsf{p} }    =
  L_\dia {^{-1/2} }
 \,\binom{ \eta\,( \tsf{r}_1+ \tsf{r}_2) ~~~~  \eta \,(\tsf{r}_1-  \tsf{r}_2) }{
  \tfrac {1}{2\,\eta} ( \tsf{p}_1+\tsf{p}_2)  ~~
  \tfrac 1 {2\,\eta}\,(\tsf{p}_1-\tsf{p}_2)},\qquad \eta=(\tfrac 1 2\, m\co\,\om_\rho){^{1/2}}.}
In general, for the two-electron problem in the magnetic field
various authors employ the Planck constant $\hbar$
(see, for example, \cite{naz}) instead of $L_\dia$.
To compare effects produced by two-body interactions with different
$M$, we have to use rescaled results. To this aim,
considering a natural relation $\a_M\,r^{-M}\sim\hbar\,\omega$,
we introduce a following definition of the parameter $L_\dia$:
\begg{ \label{PM}
   L_\dia  \equiv L_{M,s}(\a_M/\om_\rho,\eta^2), \qquad
  L_{M,s}(x,y) =  h_{M,-2\,s}\,h_{M,2}(x)\,h_{M,M}(y), \\
\intertext{where}
\label{HMK}   h_{M,k}=h_{M,k}(M),\qquad    h_{M,k}(x)= x^{k/(M+2)}\,,        }
and  $s$ is an auxiliary parameter. Next,
we introduce the dimensionless constants $(\bt,\gam)$
\begg{ \label{btm}   \a_M = \hbar\,\om_\rho\,(c/\om_\rho)^M\times \bt,  \qquad m\co =
       2\,\hbar\,\om_\rho\,c^{-2}\times \gam,  }
which yield the following relation
\begg{ L_\dia  = L_{M,s}(\a_M/\om_\rho,\eta^2) =   h_{M,-2\,s}\,\bt{^{2/(M+2)}}\,
\gam^{M/(2+M)} \times \hbar = L_{M,s}(\bt,\gam)\,\hbar.}
In particular, at $M\ss=1$ (the Coulomb interaction) we have  $\bt \equiv \a/\eps_r$
($h_{M\ss=1,-2s}\ss\equiv 1$, see Eq.(\ref{HMK})) and
\begg{
\label{P1a}
L_{\dia } = (\bt_1{^2}\,\gam)^{1/3}\,\hbar \approx
\biggl(\frac{m\co/m_{\rm e}}{ (137\,\varepsilon_r){^2} }\times
 \frac{m_{\rm e}\,c^2}{\hbar\,\om_\rho}\biggr)^{1/3}\,\hbar  \approx p
 \times ({\rm meV}/\hbar\,\om_\rho)^{1/3}\,\hbar\,,}
where $p\ss\approx 1.85$ is defined by the values of parameters $(m\co/m_{\rm e},\eps_r)$.
Thus, $L_{\dia }$ absorbs the scales related to the effective mass,
confinement energy and dielectric properties of the system.

It is instructive to caryy our analysis in terms of following variables
\begg{    \label{uvE}  u=\frac{\om_L}{\om_\rho},\qquad
      v= \frac{\om_z}{\om_\rho},  \qquad \c E_{S}\ss= |g^*|\,q_\dia\,\frac{m\co}{m_{\rm e}},
      \qquad   \om_L = \frac{e\,B}{2\,m\co},   \\
  \label{EBD}    q_\dia= \hbar/L_\dia, \qquad   E_\dia= \om_\rho\,L_\dia, \qquad
   \binom{B}{B_\dia} = 2\,m\co\,\om_\rho/e \times\binom{u}{1}, }
for the scaled cylindrical coordinates.
The quantities $(m\co,\om_\rho,L_\dia)$ establish our basic physical units, and
$(E_\dia,B_\dia)$ define the energy and magnetic strength units, respectively.
The factor $2$ in the definition of $B_\dia$ compensate the factor $\tfrac 1 2$
appearing in the definition of $\tsf{A}$:  $\tsf{A}=\tfrac 1 2\,\tsf{B}\times \tsf{r}$.
As a result, we obtain for the system Hamiltonian
\begg{ \label{ener}    H_{\ast} =(\c H_{\rm CM}+ \c H_{s,\rm rel}+ \c H_{S} )\,E_{\dia},
\intertext{where}
\label{H1}  \c H_{\rm CM} = \c H_{\rho\ast} + \c H_{z\ast}, \quad
  \c H_{s,\rm rel} = \c H_{\rho} +\c H_z + M^{s}\,(\rho^2+z^2)^{-M/2},\quad \c H_{S}= u\,\c E_S\,\hat \Sig_S,   \\
 \label{enz}  \c H_{z} \equiv   \c H_{z}(z,p_z) = \tfrac 1 2\,(p_{z}{^2}+ v^2\,{z}^2),\\
 \label{enro}  \c H_{\rho} \equiv \c H_{\rho}(\rho,p_\rho,p_\vp) =  \tfrac 1 2\,\bigl[p_{\rho}{^2}
 + {p_{\vp}}{^2}/\rho{^{2}}+(1+u^2)\,\rho{^2}\bigr] -   u\,p_{\vp},  \\
  \label{H3}    \c H_{z\ast} \equiv \c H_z(z\cov,p_{z\ast}), \qquad  \c H_{\rho\ast}=
  \c H_\rho(\rho\cov,p_{\rho\ast},p_{\vp\ast}).  }
With the aid of a transformation $\tsf{l}\to \tsf{k}= \lam^+(\tsf{l})$, where
the lists  $\tsf{l}$ and $\tsf{k}$ consist of the following variables
\begg{ \label{lk} \tsf{l} =  (\om_L,\om_\rho,\om_z,|g\co|,m\co,\a_M,M),\qquad
\tsf{k} =(u,v,\c E_S,m\co,\om_\rho,L_\dia,M), }
we  will study the Hamiltonian functions $H_\ast,H_{\dia}$ and $H\ss=H(\tsf{l},\tsf{m}_{12})$
\begg{ \label{HH}    H_{\ast}(\tsf{k}, \tsf{m}_{\ast})= H_\dia(\tsf{k}, \Upsilon(\tsf{m}_\ast)), \qquad
       H_\dia(\tsf{k},\tsf{m}_\dia) = H\circ \hat \kap^-(\tsf{k},\tsf{m}_\dia).   }
Hence, the trasformation $\hat \kap^-$ is the mapping $(\tsf{l},\tsf{m}_{12})\to (\tsf{k},\tsf{m}_\dia) $,
while the transformation  $\Upsilon = \ups\cov\times \ups$ maps the cylindrical coordinates--momenta
  onto the cartesian ones
($\ups\cov\ss\equiv \ups$); i.e.,  $\ups\cov\!: \tsf{q}\cov \to (\tsf{r}\cov,\tsf{p}\cov)$,
$\ups\!: \tsf{q} \to (\tsf{r},\tsf{p})$ where $\tsf{q}\ss\equiv (\rho,z,\vp)$.
If $M\cov$ is a phase space associated with the cylindrical coordinate system,
one has  $M\cov  \ni \tsf{m}_{\rm c} \to  \Upsilon(\tsf{m}_{\rm c})\in M_\dia$.
Hereafter, for the sake of simplicity, we drop (occurred in the function $L_{M,s}$)
the index $s$  in the notation of functions  such as  $\lam^+,\hat \kap^-,H_\dia,H_\ast$.

We have $\tsf{l}\ss \in \c L\,$, $\tsf{k}\ss\in  \c K\;$ and  $\c K\ss=\c L\ss=
 \Re \otimes \Re^6_+$, $\Re^1_+\ss=\Re_+$, $\Re^k_+\ss= \Re^{k-1}_+  \otimes \Re_+$;   hence,
   $\lam^+\in {\rm Aut}(\c L\,)$ is an automorphism of $\c L$.
The Hamiltonians $H_\dia,H_\ast$ are determined  as functions on the product of spaces
$N_\dia\ss=\c K\,\otimes M_\dia$ and $N_\ast \ss=\c K\,\otimes M_\ast$,
respectively; while the Hamiltonian $H$ is determined on the space
$N_{12}\ss=\c L\otimes M_{12}$.

Taking into account the obvious relations
\[ (\om_L,\om_z,|g\co|,\a_M)\ss= \Bigl(u\,\om_\rho,v\,\om_\rho,\tfrac{m_{\rm e}}{m\co}\,
\tfrac{L_\dia}{\hbar}\,\c E_S,\bigl( \tfrac {2\,L_\dia}{m\,\om_\rho}\bigr)^{M/2}\,\om_\rho\,L_\dia\,M^s\Bigr), \]
one obtains the inversion $\lam^-$  of the mapping $\lam^+$
\beqm{ \label{toLl}  \tsf{l}= \lam^-(\tsf{k})=
  \Biggl(k_1\,k_5,k_5,\,k_2\,k_5, \frac{m_{\rm e}}{k_4}\,\frac{k_6}{\hbar}\,k_3  ,k_5,\,
     \Bigl(\frac{2\,k_6}{k_4\,k_5}\Bigr)^{k_7/2}\,k_5\,k_6\,k_7{^s}\,,k_7 \Biggr).  }
The pair of transformations  $\{\hat \kap^-,\hat \kap^+\}$
(where $\hat \kap^+\!: N_{12}\to N_\dia$ is the inversion of $\hat \kap^-$)
are defined as
\begg{ \label{KAPPA}   \hat \kap^-(\tsf{k},\tsf{m}_\dia) = (\lam^{-}(\tsf{k}),\kap^{-}
( \lam^{-}(\tsf{k}))(\tsf{m}_\dia)), \qquad
       \hat \kap^+(\tsf{l},\tsf{m}_{12}) = (\lam^{+}(\tsf{l}),\kap^{+} ( \tsf{l})(\tsf{m}_{12} )), \\
   \label{kappa1}  \kap^\pm(\tsf{l}) =  \kap(  L_{l_2,s}{^{\pm 1 2}}
   (l_6/l_3,l_2\,l_5/2),\; 2/(3\mp 1)\times  (l_2\,l_5/2)^{\pm 1/2}).   }
Here, we use the inversion of $\kap$:  $ \kap\ord{-1}(L_\dia, \eta)\ss=
\kap(L_\dia,^{-1},\tfrac 1 {2\,\eta}) $.
The Poisson rules for the  cylindrical  coordinates  take the form
\beqm{  \label{POISSs}
   \{m^\a_{\rm c},m^\bt_{\rm c}\}_{\tsf{k},\tsf{m}_{\rm c} }  = \hat \om^{\a\bt}_{\rm c}(\tsf{k}) =
   {\rm sign}(\a)\,\del_{\a-\bt}/k_6,
    \qquad  \neg(\rho=\rho\cov\ss=0),  }
where
$(m^1_{\rm c},\ldots,m^6_{\rm c}) \equiv (\rho,z,\vp,\rho\cov,z\cov,\vp\cov)$,
$m^{-\a}\ss=p_\a,\,1\ss \le \a \ss\le 6$.
In this case we exclude the singular points at which the coordinates
$\vp,\vp\cov$ are indefinite.

Before to conclude this section there are a few remarks in order.
It is a common practice to approximate a total equilibrium
energy (described by a Hamiltonian function)
by using a finite number of terms of its Taylor series
$E_{\rm tot}=E\ord{0} + E\ord 2+...$.
For many-body problems, the first term obtained within variational approaches
is related to the classical equilibrium points upon the total energy surface of
the full Hamiltonian. This is a macroscopic part of the energy, associated very
often in quantum many-body approaches with a mean field energy (MF). The better
the macroscopic part is calculated, to a lesser degree the higher order terms
are essential. However, for finite quantum systems, quantum fluctuations
about the MF solution are quite important, which are described by
higher order terms. If the macroscopic term describes quite well
two-body correlations, the harmonic approximation associated with
the term $E\ord 2$ is good enough.

To elucidate a scale related to the term $E\ord 2$, for the sake of discussion,
let us consider the case $M=1$. The estimation of  $E\ord 2$, performed by means of
dimensionless  coordinates
$(\tsf{\bar m}\cov,\tsf{\bar m})\ss=\kap^+_{1,\eta}(\tsf{m}_1,\tsf{m}_2)$
and Eq.(\ref{EBD}), defines the microscopic (quantum) scale
\beqm{\label{DE}  E\ord 2\approx  \tfrac 1 2\,({p}^2 + \Om^2\,{q}^2)\,E_\dia  =
 \tfrac 1 2\,q_\dia\, ({\bar p}^2+ \Om^2\,{\bar q}^2)\,E_\dia   \propto \Om\, q_\dia\,E_\dia
 \propto\, \hbar\,\om_{\rho}, \nonu  }
where $\Om\sim 1$ and  $(q, p_q,q\cov,p\cov) \ss= \sqrt{q_\dia}\,(\bar q,\bar p_q,\bar q\cov,\bar p\cov)$;
and thus
$\{\bar q,\bar p_{q\pr}\}\ss= \{\bar q\cov,\bar p_{q_\ast^\prime} \} =\hbar^{-1}\,\del_{q q\pr}$.

In our model, the macroscopic part can be estimated by means of the classical approach,
omitting the contribution of the spin interaction in the Hamiltonian (\ref{ener}).
Finally, taking into account the contribution of quantum oscillations,
we will include the contribution of the Zeeman interaction.

It appeares that the quantity $q_\dia$  (see Eqs.(\ref{P1a}),(\ref{EBD}))
characterizes the strength of the quantum effects over the classical ones.
Indeed, at $q_\dia \ss\sim (\hbar\,\om_\rho/{\rm meV})^{1/3} \to 0$
the contributions of the second and higher order terms in the Taylor expansion of
the total energy are much smaller then a principle (macroscopic)
part of the energy found by means of the minimization of the
Hamiltonian function $H$.

\subsection{ \label{SYM} Symmetries}

According to the  decomposition of a phase space $M_\dia\ss=M_{\rm CM}\otimes M_{\rm rel}$,
 the  group  of canonical symmetries $G_{M_\dia}$
factorizes onto the direct group product:  $G_{M_\dia} \ss=G_{\rm CM} \times G_{\rm rel}$.
Here  $G_{\rm CM}\ss=U(2)\times U(1)$,  where $U(2)$ acts on  the complex vectors
$  [p_{x\ast}+\,i\,m\co\,\om_\rho\,\sqrt{1+u^2}\,x\cov ]\,\vec e_1 + [ p_{y\cov}+\,i\,m\co\,
\om_\rho\,\sqrt{1+u^2}\,y\cov ]\,\vec e_2 $,
while  $U_1$  acts on a complex vector  $(p_{z\ast}+i\,m\co\,\om_z\,z\cov)$.
The transformation  group $G_{\rm rel}={\rm SO}_{\vec e_z} (2)\times \Pi$
defines the symmetries of $H_{\rm rel}$. Here
 $ {\rm SO}_{\vec a}(2) \ss= \bigcup_{\psi\ss=0}^{2\pi}R_{\vec a}(\psi)$,
$R_{\vec a}(\psi)$ is a rotation  around vector $\vec a$;
$\Pi\ss=\{E,\c P_{\rm in}\}$ is the discrete group, where $E$ is a neutral element, while
$\c P_{\rm in}\cdot (\tsf{r},\tsf{p})= (-\tsf{r},-\tsf{p})$ is the inner parity operator.

The analysis of symmetries of a function $H$  (or $H_{\dia}$) is desirable to
study, considering the Hamiltonian as  a function on the space
$N$ where  $N\ss= N_{12}$ or $N\ss=N_\dia$
(see Eq.(\ref{HH})). Consequently, the concept of symmetry group  is more
convenient to formulate,  studying  the group of automorphisms  ${\rm Aut}(N)$
of a space $N\ss= \c L \otimes M$, constrained from the requirements of invariance
of a symplectic two--form
\beqm{\label{OmNG}   \Om_N(\tsf{l},\tsf{m}) \ss= \Sig_{\a>\bt} \,\om_{\a\bt}(\tsf{l},\tsf{m})\,dm^a\we dm^\bt.  }
In general (see, for example, a textbook \cite{Arnold}),
the simplecticity of $\Om_N$ is assured by the pair of conditions:
$(a)$  ${\rm det}\,\om(\tsf{l},\tsf{m})  \neq 0$ and
 $(b)$  $\om_{\a\bt,\gam}+ \om_{\bt\gam,\a}+ \om_{\gam\a,\bt}\ss=0$ for all $\a,\bt,\gam$ and
  $f_{,\a}(\tsf{l},\tsf{m}) \ss\equiv (\pl_{m^\a}\,f)(\tsf{l},\tsf{m})$.

Note that $\om_{\a\bt}$ are elements of a  covariant skew symmetric tensor
$\om\ss\equiv \om(\tsf{l},\tsf{m})$, fulfilling  the identity
$\om \cdot \hat \om\ss=1 $, where
  $\hat \om^{\a\bt}\ss\equiv  \hat \om^{\a\bt}(\tsf{l},\tsf{m})\ss=
  \{m^\a,m^\bt\}_{(\tsf{l},\tsf{m})}$ are elements of the contravariant one.
Hence, if $\tsf{m}\ss=\tsf{m}_{12}\ss=
(m^1_{12},\ldots,m^6_{12},m^{-1}_{12},\ldots,m^{-6}_{12})$ is
a set of the canonical coordinates and
  $\hat \om \ss= \hat \om_{12}$, then
 $\hat \om^{\a\bt}_{12}\ss= -\om_{12,\a\bt}\ss= {\rm sign}(\a)\,\del_{\a-\bt}$ for
 $m^\a_{12}\in \{x_1,y_1,z_1,x_2,y_2,z_2\}$, $m^{-\a}_{12}\ss=p_{q^\a}$. In this case
 $\Om_{N_{12}}$ takes  the standard (canonical) form:   $\Om_{N_{12}}(\tsf{l},\tsf{m}_{12}) =
 \Om_{M_{12}}(\tsf{m}_{12})\ss=\Sig_{i\ss=1}^2\, \Sig_\a\,dp^\a_i\we dr^\a_i$.
Due to validity of the relation $dp_{a1}\we dr_{\a 1}+ dp_{\a 2}\we dr_{\a 2}\ss=
L_{s}(\tsf{l})\,(dp_\a\we dr_{\a} + dp_{\a \ast}\we dr_{\a \ast})$ one finds
\beqm{ \label{OmN}   \Om_{N_\dia}(\tsf{k},\tsf{m}_\dia) =
k_6\;\SUM{\a}{}\,(dp_\a \we dr_\a + dp_{\a \ast}\we dr_{\a \ast}), \qquad k_6\ss\equiv L_\dia.  }
Here $\Om_{N_\dia} = \hat \kap^{-\ast}\,\Om_{N_{12}}$ is a pull back of $\Om_{N_{12}}$ induced
through the  mapping $\hat \kap^-$: $N_\dia \mapsto N_{12}$.
The application of identity  $\om_\dia\cdot \hat \om_\dia\ss=\tsf{1}$
for $(\om_\dia)_{\a\bt} \ss\equiv - {\rm sign}(\a)\,k_6\,\del_{\a-\bt} $
(see Eq.(\ref{OmNG},\ref{OmN})) establishes the following Poisson brackets on $M_\dia$:
\beqm{ \label{POISS}   \{m^\a_\dia,m^\bt_\dia\}_{(\tsf k,\tsf{m}_\dia)} \ss=
  \hat \om^{\a\bt}_\dia(\tsf{k},\tsf{m}) ={\rm sign}(\a)\,\del_{\a -\bt}/k_6,   }
where  $(m^1_\dia,\ldots,m^6_\dia) \equiv(x,y,z,x\cov,y\cov,z\cov)$, $m^{-\a}_\dia \equiv p_\a$, $M_\dia$.
If $\Om_{N_i}(\tsf{k}_i,\tsf{m}_i)=
 \Sig_{\a>\bt}\,\om^i_{\a\bt} (\tsf{k}_i,\tsf{m}_i)\\ \times dm^\a_i \we dm^\bt_i$  then
\begg{ \label{omg}  \om^2_{\a\bt} (\tsf{k}_2,\tsf{m}_2)\ss=
  \Sig_{\gam \del}\,\om^1_{\gam \del}\circ \Gam_{\tsf{k}_2}(\tsf{m}_2) \times
  \Gam^\gam_{\tsf{k}_2,\a}(\tsf{m}_2)\,\Gam^\del_{\tsf{k_2},\bt}(\tsf{m}_2).   }
Here $\Gam$: $N_2\to N_1 \ni (\tsf{k}_1,\tsf{m}_1) \ss=\Gam_{\tsf{k}_2}(\tsf{m}_2)$,
and the rule (\ref{omg})  results from   the condition $\Om_{N_2} \ss= \Gam\co \Om_{N_1}$.
In particular, applying (\ref{omg})   for $\Gam= ({\rm id}_K, \Upsilon)$,
$N_1\ss=N_\dia$, $N_2\ss=\c K\;\otimes M_\ast$,
$\om^1\ss=\om_\dia$, $\om^2\ss=\om_\ast$,
 one finds:   $\om_{\ast,\a\bt}(\tsf{k},\tsf{m}_\ast)\ss\equiv\om^2_{\a\bt}(\tsf{k},m\cov) \ss
 = \om_{\dia, \a \bt}(\tsf{k}) \ss= -{\rm sign}(\a)\,\del_{a-\bt}/k_6 \Rightarrow \hat \om_\ast\ss=\hat \om_\dia$.
Thus,  the right hand sides of Eqs.(\ref{POISSs}) and (\ref{POISS})   are identical.
The formula (\ref{omg})  provides  a simple tool for study the group of symmetry.

In particular, let us consider Eq.(\ref{omg}) at the conditions  $N_1\ss=N_2\ss=N$, $\Gam\ss=g$,
$\om_{\a\bt}(\tsf{k},\tsf{m})  = \om_{\a\bt}(\tsf{k})$   (see, Eq.(\ref{OmNG})),
and let $\om^2\ss=\om^1 \ss=\om$, where $\om_{\a\bt}(\tsf{k},\tsf{m})\ss \equiv \om_{\a\bt}(\tsf{k})$;
hence,
\begg{
  \label{om}   \Sig_{\gam \del}\,\om_{\gam \del}\circ \check g(\tsf{k})\times
  \hat g^\gam_{\a}(\tsf{k},\tsf{m})\,\hat g^\del_{\bt}(\tsf{k},\tsf{m})\ss= \om_{\a\bt}(\tsf{k}).}
This formula expresses the condition of the invariance for $\Om_N$: $g\co \Om_N\ss=\Om_N$ where
$g \in G_\ltimes$,
\beqm{ \label{Gsem}     G_\ltimes  =  {\rm Aut}(\c K)  \ltimes {\rm Aut}(M), }
induced  through a factorization of the element $g$: $g\ss= \check g\circ \hat g$ where
$\check g\in {\rm Aut}(K)$, $\hat g\in {\rm Aut}(M)$.
Assumptions  that $\check g,\hat g$ are elements of  a semidirect group product $G_\ltimes$
had been applied in Eq.(\ref{om}). These elements  are given in the following forms:
$\check g(\tsf{k},\tsf{m}) \ss=(\check g(\tsf{k}),\tsf{m})$,
 $\hat  g(\tsf{k},\tsf{m}) \ss= (\tsf{k},\hat g(\tsf{k},\tsf{m}))$.
 It results in the following rules of composition (multiplication) of group elements:
$(g_1\circ g_2)(\tsf{k},\tsf{m}  )\ss= (\check g_1\circ \check g_2(\tsf{k}),
\hat g_1\circ g_2(\tsf{k},\tsf{m}))$.

The group $ {\rm Aut}(M)$ is the normal subgroup of $G_\ltimes$:
$(g_1 \circ \hat g\circ g_1{^{-1}})(\tsf{k},\tsf{m})\ss=
(\tsf{k},(\check g_1\circ g\circ g_1{^{-1}})(\tsf{k},\tsf{m})) \in {\rm Aut}(M)$,
while ${\rm Aut}(\c K\,)$ is a factor group.

For the chain $G_F \subset G \subset G_\ltimes$ one has:
\llist{
\itt{1} $G$  is a group of automorphisms of $(N,\Om_N)$, i.e.  $g\in G$ obeys the condition (\ref{om});
\itt{2} $g\in G_F$  is a symmetry of $F$: $g\in G$ and $F\circ g(\tsf{k},\tsf{m})\ss= F(\tsf{k},\tsf{m})$.}
In general, neither the factor  $\check g$, nor $\hat g$  are  elements of  $G$  (or  $G_F$).

Let us consider the following two--dimensional transformations of the space $N_{12}$:
\begg{ \label{gab}   g_{a,b}(\tsf{l},\tsf{m}_{12}) = (\tsf{l}_{a,b}, g_a(\tsf{m}_2)), \qquad
   g_{a} (\tsf{r},\tsf{p}) = (a^{-1} \tsf{r},a\,\tsf{p}), \\
  \label{gab1}      \tsf{l}_{a,b} =
  (\om_L,\om_z,(a\,b)^{-2}\,|g\co|,a^2\,m\co,\om_\rho, a^2\,(a\,b)^{-M-2}\,\a_M, M). }
Here,  $g_a$ is an element of a group of linear  symplectic transformations
$g_b \ss\in {\rm Sp}(12,\Re)$.
The transformation $\tsf{l}\to \tsf{l}_{a,b}$ has also  the linear and diagonal
form: $l_{a,b}^i \ss= f_i(a,b)\,l^i$.
To elucidate a physical interpretation of these transformations,
all elements of list $\tsf{l}_{a,b}$   in the right hand side of Eq (\ref{gab1})
have been replaced by their   original physical values (see Eq.(\ref{lk})).

The physical interpretation of transformations  $g_{a,b}$
is  much transparent, if one replaces $g_{a,b}$ by  their  images  $g_{\dia,a,b}$.
Thus, let   $g\in {\rm Aut}(\c L\;)\ltimes {\rm Aut}(M_{12})$ and
$g_\dia\in {\rm Aut}(\c K\;)\ltimes {\rm Aut}(M_\dia)$,
then the adjoint transformation
\beqm{ \label{Adp}   {\rm Ad}_{\hat \kap^+}\!:  \quad   g   \to
g_\dia =    \hat \kap^+ \circ  g   \circ \hat \kap^-, }
establishes the  group of homeomorphisms
$ {\rm Aut}(\c L\;)\ltimes {\rm Aut}(M_{12}) \mapsto  {\rm Aut}(\c K\;)\ltimes {\rm Aut}(M_\dia)$.
With the aid of Eqs. (\ref{PM},\ref{HMK},\ref{lk}--\ref{kappa1},\ref{gab}, \ref{Adp}) one   finds:
\begg{\label{ST}   g_{\dia,a,b}(\tsf{k},\tsf{m}_\dia) = {\rm Ad}_{\hat \kap^+}(g_{a,b})(\tsf{k},\tsf{m}_\dia)=
        ((u,v,\c E_S,a^2\,m\co,\om_\rho,b^{-2}\,L_\dia,M),b\,\tsf{m}_\dia). }

The validity of the condition  Eq.(\ref{om}) in the case
$g_{\dia,a,b}(\tsf{k},\tsf{m}_\dia)$  and $\om_{\a\bt}(\tsf{k}) \ss= \om_\dia(\tsf{k})$ is trivial.
The obvious  relation $(\pl_{k_4  =   m\co}\,H_{\dia,s})(\tsf{k},\tsf{m}_\dia)\ss=0$
proves  that the group $G_4\ss=\{g_{\dia,a,1}:a\in \Re\}$  establishes  the symmetries of $H_\dia$.

The physical interpretation of elements $g_{\dia,1,b}$ is determined in the limit
$\a_M\ss=0$, i.e., when $q_\dia\ss=\infty$. In this case, the dimensionless Hamiltonian
$\c H_\dia \ss=\c H_\dia(\tsf{k},\tsf{m})$ is a homogenous function
 of $\tsf{m}$:
$\c H_\dia( \check g_{\dia,a,b}(\tsf{k}), \hat g_{\dia,a,b}(\tsf{k},\tsf{m}))=
\c H\,(\tsf{k},b\,\tsf{m}_\dia)\ss= b^2\,\c H\,(\tsf{k},\tsf{m})$ and
$E_\dia( \check g_{\dia,\a,b} (\tsf{k} ) )  \ss= b^{-2}\,E_\dia(\tsf{k})$, where
$E_\dia(\tsf{k})\ss=k^5\,k^6\ss\equiv \om_\rho\,L_\dia$;
hence $H_\dia \ss= \c H_\dia\times E_\dia$ is $g_{\dia,1b}$ invariant.
We conclude that the elements $\hat g_{\dia, a,b}\!: a \in \Re, b\in \Re $
form the asymptotic symmetry group for the Hamiltonian $H_\dia$.

Thus, we proved that a symmetry group of the Hamiltonian system
$(N_\dia,\Om_{N_\dia},H_\dia)$ is established by the direct group product:
\beqm{  \label{GNd}     G_{N_\dia}  =  G_{4}   \times G_{M_\dia}
\subset {\rm Aut}(\c K\,) \times {\rm Sp}(M_\dia), }
where ${\rm Sp}(M_\dia) \subset {\rm Aut}(M_\dia)$ results  from the application
of condition (\ref{om}) for $\om\ss=\om_\dia$.

The physical interpretation of  group $G_4$ follows from  the  invariance of
parameters $\om_L,\,\om_\rho,u\ss=\om_L/\om_\rho$ and
the transformation rule for $m\co$: $m\co_a\ss=g_{\dia,a,1}\cdot m\co \ss= a^2\,m\co$
(see Eq.(\ref{ST})).
Using these rules,  we obtain
\beqm{ \label{Ba} B_a/u   =   (B/u)_a  \equiv B_{\dia\,a} =  (2\,m\co\,\om_{\rho}/e)_a=
a^2 \times 2\,m\co\,\om_{\rho}/e\ss= a^2\,B/u,  }
where $x_a  \ss\equiv \hat g_{a1}\cdot x_a$.  Thus, we can propose
two different physical  interpretations of symmetries $G_4$:
\llist{
\itt{\rm I} if $m\co$ is an unknown parameter,  the effective mass  is established
by the action of the group element $g_{\dia,a,1}$:
    $m\co\equiv m_0\,a^2$, where  $m_0$ is a constant; while
  the physical value of parameter  $a$ is  determined from the experimental data:
   $u \to B\ss= u\times 2\,(m_0\,a^2)\,\om_\rho/e$,
\itt{\rm II} if effective masses of different experiments are known,
    the group  $G_4$  conjugates states of different physical  systems; thus, enabling
    to predict, for example, results of one experiment from those of the other one. }
\section{\label{SB} Families of equilibrium states}
At fixed values of the integrals of motion  $p_{\vp}\ss=p_0$, $p_{\vp\cov}\ss=p_{0\ast}$
one is faced with   a reduced   Hamiltonian
dynamics of the rest of the canonical variables:
$((\rho,z,\rho\cov,z\cov), (p_\rho,p_z,p_{\rho\ast},p_{z\ast}))$.
In other words, we have to solve the minimization problem for the reduced Hamiltonian dynamics
with respect to the canonical variables of the reduced phase space. One obtains
\def\dub#1#2#3{\raise #3\hbox{$#3$}\box{#2} }
\begg{  \label{dH}   d H_{\ast}  = d E = (\dot \vp\,dp_{\vp} +\dot \vp\cov\,dp_{\vp\cov})\,E_\dia  \Rightarrow
  \binom{\dot \vp}{\dot \vp\cov}  = \binom{\pl_{p_{\vp}}}{ \pl_{p_{\vp\cov}}} \,H_{\ast,s} =
  \binom{u_{\tsf{rot}}-u}{u_{\tsf{rot}\cov} -u}\,E_\dia ,  \\
 \label{uuc}        u_{\tsf{rot}} = {\rho}{^{\,-2}}\,p_\vp, \qquad  u_{\tsf{rot}\cov}
 ={\rho\cov}{^{\,-2}}\,p_{\vp\cov}. }
The equilibrium points are determined by means of the following equations of motion
\begg{ \label{EXTRE}  0=\dot{q} = L_\dia{^{-1}}\,\pl_{p_q}\,H_{\ast} = \om_\rho\,\pl_{p_q}\,\c H_{\ast},\qquad
  0=\dot {p}_q=-L_\dia{^{-1}}\,\pl_{q}\,H_{\dia} =-\om_\rho\,\pl_{q}\,\c H_{\ast},
\intertext{where $q\ss=z,\rho,\rho\cov,z\cov$.
As a result, we have six elementary conditions}
\label{incon}    0 = p_z=p_\rho=p_{\rho\cov}=p_{z\cov}= z\cov=0 , \\
\label{rhocov}  \rho\cov{^2}  =(1+u^2)^{-1/2}\,  |p_{\vp\cov}|.}
These conditions provide the definition of the  centre-of-mass energy
$\c E_{\rm CM}$  in $E_\dia$ unit.
\beqm {  \label{cm}  \c E_{\rm CM}\ss= (1+u^2)^{1/2}\,|p_{\vp\cov}| - u\,p_{\vp\cov}.}
Two nontrivial requirements are obtained for the relative motion which depends on
$(\rho,z)$ coordinates. In particular, for $z$ coordinate Eqs.(\ref{EXTRE}) lead  to the condition
\begg{
        \label{con1}  0= \pl_{z}\,\c H_{s,\rm rel} =  z\times (v^2- M^{s+1}\, r^{-M-2}).}
We recall that  $s$ is the additional parameter, which is  not fixed yet.
Hereafter, in order to simplify analytical expressions at
the equilibrium values of $r$, we take $s\ss=-1$.

The condition (\ref{con1}) is fulfilled at: $z=0$ and
\beqm{   \label{zero}
        z\neq 0\; \Longrightarrow \; r= r_{\tsf{A}}(v,M) =  h_{M,-2}(v),}
where $h_{M,k}(x)$ is defined by Eq.(\ref{HMK}).
For $\rho$ coordinate we obtain the following condition
\begg{  \label{Hrho}   0 = \pl_{\rho}\,\c H_{s=-1,\rm rel} = (1+u^2)\,\rho^2 -
(p_\vp/\rho)^2 - \rho^2\,r^{-(M+2)}.   }
At $z=0$, Eq.(\ref{Hrho}) defines the family of {\it symmetric} states
(due to the reflection symmetry: $(\rho,z)\to(\rho,-z)$).
For $z\neq 0$, Eqs.(\ref{zero},\ref{Hrho}) yield the solutions for the variable $\rho,z$:
 $(\rho,z)\ss=(\rho_{\tsf{A}},\pm z_{\tsf{A}})$ where
\begg{  \label{Aeq}  \{\rho_\tsf{A},z_\tsf{A}\}  = d^{-1/2}(u,v)\times \bigg\{ |p_\vp|^{1/2},
  \,\sqrt{d(u,v)\,h_{M,-2}(v^2)  - |p_\vp|} \bigg\},  }
further studied   as  $g_{\tsf{A},u,v,M}(p_\vp)$ functions ($g\ss=\rho,z$). Here, we also
introduced the notation
\beqm{      \label{dd}  d(u,v)= \sqrt{1+u^2-v^2}. }
\begin{figure}[!t]
\centering   \includegraphics[scale=.9]{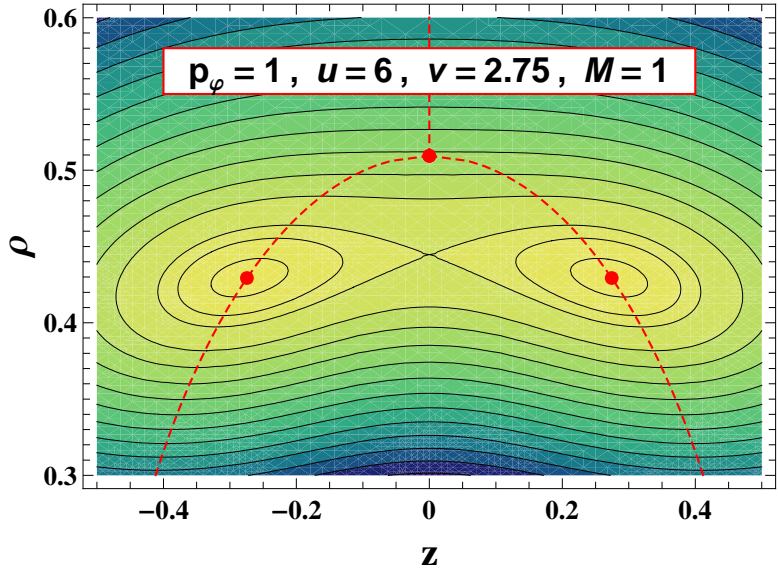}
 \caption[$(Z,\rho)$ projection of Energy contours] {   Energy surface defined
 for $M=1$ (the Coulomb potential) at $p_z=p_\rho=0$ and $p_\vp=1$.
 The energy is calculated for $u=\omega_L/\omega_\rho=6.0$ and for
 the $3$D system with $v=\omega_z/\omega_\rho=2.75$.
 There are two distinct minima at $z\neq 0$. The dashed lines indicate the
 possible pathes along which the system moves to the symmetric minimum
 $z_\tsf{S}=0$ with   the increase of the angular momentum. }
 \label{fig1}
\end{figure}
Thus, there are two families of equilibrium states for the relative motion:
\llist{
\itt{a} {\em asymmetric} states ($\tsf{A}$): $(\rho,z)\ss=(\rho_\tsf{A},z_\tsf{A})$,  $d^{\,2}(u,v)>0$\,;
\itt{b} {\em symmetric}  states ($\tsf{S}$): $z\ss=z_{\tsf{S}}\ss=0$.  }
Note that the condition $d^{\,2}(u,v)>0$  restricts the lower limit of the magnetic
field for the existence of the {\em asymmetric} states. These states could exist
only for the condition
\begg{   \label{fizsol}   d^{\,2}(u,v)>0 \Rightarrow \omega_\rho{^{2}}+\omega_L{^{2}}\ge\omega_z{^{2}}.    }
For the sake of illustration, we calculate the total classical energy for $M=1$
(the Coulomb potential), defined 
by Eqs.(\ref{ener},\ref{enz},\ref{enro}) for fixed values of parameters
(see Fig.\ref{fig1}). Two asymmetric minima $z\neq0$ of the Hamiltonian function
(\ref{ener}) are exhibited on the energy surface for a given value of the
angular momenta ($p_\vp=1$) at the fixed values of the magnetic field ($u$)
and the system (QD) size ($v$). For the fixed parameters the increase of the angular momentum
value transforms two asymmetric minima   $(\rho,z)\ss=(\rho_\tsf{A},\pm z_\tsf{A})$ to the
symmetric one $(\rho,z)\ss=(\rho_\tsf{S},0)$
which moves along the vertical line.
We return to this point in next Section.
\subsection{Asymmetric states}
Let us focus on the family of equilibrium solutions for the asymmetric states.
Eq.(\ref{Aeq}) determines   the  equilibrium energy of the relative motion
\begg{
\label{en1}      \c E_{\,\tsf{A}, u,v,M}(p_\vp) =  \tfrac 1 {2}\,(1 +\tfrac{2}M)\,h_{M,2\,M}(v) +
     d(u,v)\times|p_\vp|  -u\,p_\vp.  }
Here, the relative energy  $\c E_{\,\tsf{A}, u,v,M}(x)$  as well as
the centre-of-mass energy $\c E_{\,\rm CM}$ (see, Eq.(\ref{cm}))
 is defined in $E_\dia$ units.
The equilibrium states create the energy hyper-surface in the three-dimensional
space of physical external parameters $(u,v,p_\vp)$. Evidently, this surface
is bounded by the families of $\it symmetric$ states. Our aim is
to find a range of the parameters which determine the
{\it asymmetric} states on the energy hypersurface of extreme states $(\c E,u,v,p_\vp)$.

To proceed we introduce the following function
\beqm{  \label{gm}     G_M(p_\vp,u,v)=   h_{M,8}(v)\,{p_\vp}{^2}-u^2+ v^2.  }
With the aid of this function let us consider the ratio
\beqm{  \label{max}      \rho_\tsf{A}^4/r_\tsf{A}{^4}  =  (G_M(p_\vp,u,v)+u^2-v^2)/d^{\,2}(u,v).  }
Evidently, the condition  $ G_M(p_\vp, u,v)=1$  yields the solution $z_\tsf{A}=0$.
\ndef{ \label{TheoA}   Mapping $\tsf{A}_{u,v}: p_\vp \mapsto \tsf{q}_{u,v,M}(p_\vp)$,
where $\tsf{q}_{u,v,M}=(r,r\cov)$ and $r:=(\rho,z,p_\rho,p_z)$,
we call $\tsf{A}_{u,v}$ band.
      The ranges of the physical parameters $u\, ($for convenience,
      we consider the positive magnetic field $u>0 $; see below$)$
      and $p_\vp$ are  obtained from the inequality:
\beqm{  \label{SIG}    G_M(p_\vp,u,v) \le 1.}
Points of the set $\Sig_\tsf{A}(u,v)=\{(u,v,p_\vp),\, G_M(p_\vp, u,v)=1\}$ we
call  maximal $\tsf{A}_{u,v}$ states.}
The maximal $\tsf{A}_{u,v}$ states are points of the intersection between $\tsf{A}$
and $\tsf{S}$ sets: $\Sig_{\tsf{A}}\ss=\tsf{S}\cap \tsf{A}= \pl \tsf{A}$, i.e.,
the set $\Sig_\tsf{A}$ closes  the family of  $\tsf{A}$ states.

The  inequality  (\ref{SIG}) can be resolved with respect to $u$ or to $p_\vp$:
\begg{
u_{\rm crit}(v,p_\vp,M)   \leq u \;\;  \we \;\;    |p_\vp| <p_{u,M}(v),
\intertext{where}
  \label{ucr}      u_{\rm  crit}(v,p_\vp,M)=\sqrt{h_{M,8}(v)\,{p_\vp}{^2}-1+ v^2},  \\
       \label{pcr}       p_{u,M}(v)       =d(u,v)\,h_{M,-4}(v). }
The transition point from the family of $\tsf{S}$ states to the family  $\tsf{A}$ states
 signals on the {\it spontaneous symmetry breaking}
with respect to the reflection $(\rho,z)\rightarrow (\rho,-z)$ at a fixed value
of $p_{u,M}$.
We recall that, in general, the spontaneous symmetry breaking is
associated with the symmetry breaking of the system's ground state, although
the symmetries of the Hamiltonian hold true (cf \cite{PR}).
We are faced with the spontaneous breaking of the
inner parity symmetry $\c P_{\rm in}$ at the preserved integral of motion
$p_\vp=${\sl const}. Thus, there is a coexistence of two families of states which
we associate with two phases at a
fixed value $p_\vp$, and the set  $\Sig_\tsf{A}(u,v)$ determines
the unstable $\tsf{S}$ states.
The question arises: what kind of states ($\tsf{A}$ or $\tsf{S})$
describes the ground and excited states  in the manifold $(u,v,p_\vp)$,
where the parameters $(u,v)$ are external parameters of the system ?
Below we aim to define the family of stable $\tsf{S}$ states and to
illuminate the question about the equilibrium states of the system.
\subsection{Symmetric states}
From the evident relation for the Hamiltonian of the relative motion
(see Eqs.(\ref{ener},\ref{enz})) we obtain for the symmetric states $\tsf{S}$
\beqm{
\label{indv} \pl_v {\c H}= z^2\bigl{|}_{z_\tsf{S}=0}=0.}
It results in that the equilibrium values of the variable $\rho\ss=\rho_\tsf{S}$ and the equilibrium
energy $\c E\ss=\c E_{\tsf{S}}$  have to be  independent functions of the external parameter $v$:
\beqm{   \pl_{v} \rho_\tsf{S}\ss=\pl_v \c E_{\tsf{S}}\ss=0.   }
It means that for the maximal states $\tsf{A}_{u,v}$ the
{\em Definition}  \ref{TheoA} serves as a constraint for the definition
of values $v=v\cov$ as a function of the $p_\vp,u$ :
\beqm{ \label{vGM}  v=v\cov, \qquad G_M(p_\vp,u,v\cov)\ss=1, \nonu }
such that the transition from the family of $\tsf{S}$ states
to the family $\tsf{A}_{u,v}$ can be interpreted as the tendency of the
$z$-vibration frequency  for $\tsf{S}$ state to approach zero.

All remaining $\tsf{S}$ states are found applying to  the elements  of sets
$\Sig_{\tsf{A}}(u,v\cov)$ the transformation
$T_\a\!: (u,v\cov,p_\vp=p_{u,M}(v\cov),\rho_{\tsf{S}},z_{\tsf{S}})\to(u,v=
\a\,v\cov,p_\vp,\rho_{\tsf{S}},z_\tsf{S}),\,1\neq \a\in \Re_+$.
We obtain a new set of independent variables $(u,v,v\cov)$
instead of the old one $(u,v,p_\vp)$.  As a result, the above consideration
enables to one to obtain  the $\tsf{S}$ family  in  the following way:
%
%
\ndef{ \label{SDEF}  The space of $\tsf{S}$ states is constructed from $A_{u,v}$
states considering a three--dimensional manifold$:$}
\vskip -5mm
\beqm{ \label{sps}   \tsf{S} =  \{(p_\vp,u,v,v\cov),\;G_M(p_\vp,u,v\cov)\ss=1,\,
v \in \Re_+,\,v\cov \in \Re_+\},   }
We find immediately  $(p_\vp,u,v,v\cov)\in \tsf{A}_{u,v} \cap \tsf{S} \Leftrightarrow v\cov\ss=v$; hence,
the mapping $v\cov \to p_\vp$,
\beqm{ \label{pxM}   p_\vp = \eps_\vp\, p_{u,M}(v\cov), \quad \eps_\vp=\pm 1.  }
provides the states consistent with {\em Definition} \ref{SDEF}. Namely,
$G_M(\eps_\vp\,p_{u,M}(v\cov), u,v\cov)=1$  for $p_{u,M}$  determined by Eq.(\ref{pcr})
turns to be the  identity  relation.  Consequently, the expressions
for $\tsf{S}$  functions  can be found employing the expressions for
$\tsf{A}$ functions with the aid of the rule
\beqm{ \label{fS}   f_{\tsf{S}}(u,v,p_\vp,M)\rightarrow f_{\tsf{A}}(u,v\cov,p_{u,M}(v\cov),M),
       \qquad f = \rho,\c E. }
In particular, Eqs.(\ref{Aeq},\ref{pxM}) determine an
equilibrium value $\rho$: $\rho\ss=\rho_S(v\cov,M)$;  hence,
\begg{ \label{uvrep}   \rho_S(v\cov,M)= r_\tsf{A}(p_{u,M}(v\cov),u,v\cov,M) =
            h_{M,-2}(v\cov).     }
From   Eqs.(\ref{en1},\ref{pcr},\ref{fS}), in the same manner, one obtains
the energy of relative motion
\begg{   \label{Euv}    \c E_{\ast \tsf{S},u,M,\eps}(v\cov) = \tfrac 1 2\,
(1+\tfrac 2 M)\,h_{M,2\,M}(v\cov)+ h_{M,-4}(v\cov)\,   d\,(d  - \eps\,|u|)\,,}
where $d\ss\equiv d(u,v\cov)$.

The reflection $(p_\vp,u)\mapsto (-p_\vp,-u)$ is the Hamiltonian symmetry.
Therefore, it is convenient to choose a positive magnetic field ($u\ge0)$, while
to analyze both negative and positive values for $p_\vp$.
Here $\eps\ss= \eps_\vp\,\eps_u=\eps_\vp$ due to our choice $\eps_u\ss={\rm sign}\,u=1$.

Evidently, the equilibrium values  $(\rho,p_\vp,\c E\,)_{\,\tsf{S}}$ do not depend on
the variable $v$.  The set of equations describing the equilibrium $\tsf{S}$ states
with the aid of the variable  $v\cov$ we name the
$(u,v,v\cov,\eps)$- (or shortly) $v\cov$-parameterizations of $\tsf{S}$  states.

It is useful to exclude the parameter  $v\cov $.
In virtue of definition of $v\cov$ as the value $v$ of maximal $\tsf{A}_{u,v}$ states
(see  {\em Definition} \ref{TheoA} and {\em Definition} \ref{SDEF}) and  the
definition of $G_M$ in    Eqs.(\ref{gm}), one obtains the following equation
for the variable $v\cov$
\begg{   \label{vcp}
     0=-p_\vp{^2}+ h_{M,4}\, (1 + u^2)\, Z^4 - h_{M,4}\, Z^{2 - M},    \qquad Z = h_{M,-2}(v\cov), }
as  a function  of fixed parameters $(p_\vp,u,M)$. In particular, for $M=1$
there is a single real solution
\begg{ \label{vcov} v\cov(p_\vp,u,M=1) = Z^{-3/2}(p_\vp,u), \quad Z(p_\vp,u) =
\frac{\sqrt{s}}{2} + \frac 1 2\,\sqrt{-s+\frac{2}{\sqrt{s}\,(1+u^2)}},\\
s= -\frac{4\,2^{1/3}\,p_\vp{^2}}{Q}  + \frac{Q}{3\,2^{1/3}\,(1 + u^2)}, \\
 Q = \Bigl(27 + 27\,u^2 + \sqrt{729\,(1 + u^2)^2 + 6912\,p_\vp{^6}\,(1 + u^2)^3}\Bigr)^{1/3}. }
As a result,  for the equilibrium $\tsf{S}$ states one obtains
\begg{  \rho_{\,\tsf{S}}\vert_{M=1}  =v^{-2/3}_\ast(p_\vp,u,1)\ss=Z(p_\vp,u).}
\begin{figure}[!t]
 \centering{   \includegraphics[scale=.8]{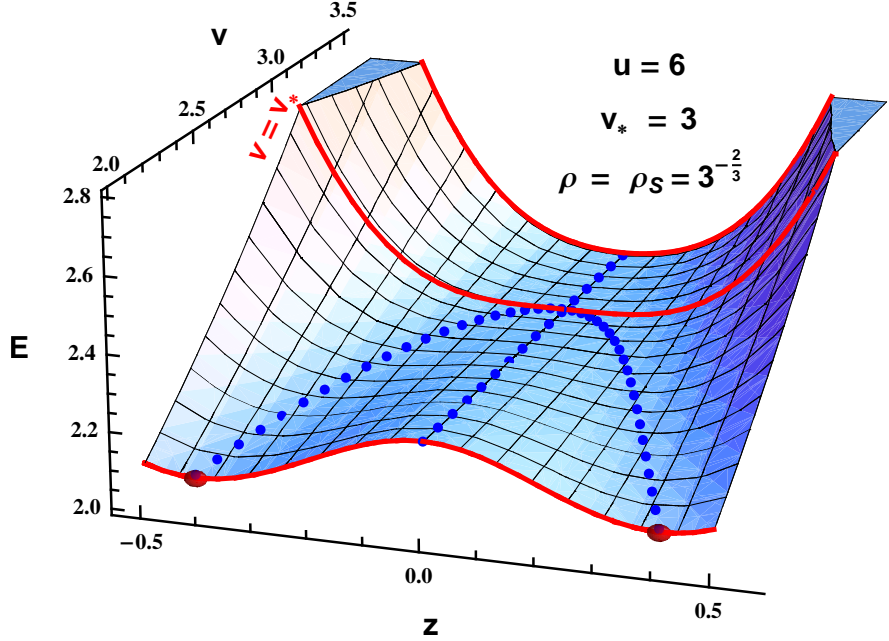} }
 \caption[The energy surface]{The surface of energy of relative motion
 studied by means of the $v\cov$  parametrization for $M\ss=1$,
 $\rho\ss=\rho_{\tsf{S}}(3,1)$, and $u=\omega_L/\omega_\rho=6.$
 The energy is given in $E_\dia$ unit. }
 \label{fig2}
\end{figure}
The  parametrization $(u,v,v\cov,\eps)$ or
$v\cov$--parametrization  is a key element which enables to one to elucidate
the shape transition phenomenon. Assuming $M\ss=1$, $v\cov\ss=3$ and
$\rho\ss=\rho_{ \tsf{S} }(3,1)$ (see Eq.(\ref{uvrep})),
with the aid Eqs.(\ref{H1}-\ref{enro})  we consider the energy  surface
$\c E= \c H_{s,\rm rel}$ at $p_\rho=p_z=0$
and $(\rho,p_\vp)\ss= (\rho_\tsf{S}(v\cov,M),p_{u,M}(v\cov))$  (see Fig.\ref{fig2})
as function of $(v,z)$ variables. The magnetic field strength is chosen as
$u=\omega_L/\omega_\rho=6$. Accordingly with  Eqs.(\ref{indv}),(\ref{Euv})),
along the line $z\ss=0$, the  energy does not depends on $v$ (dotted line).
The curve $v\ss=v\cov\ss=3$ divides a plane $(z,v)$ on two domains.
For each  section  $v$ such that $v<v\cov$ the condition
$z\ss=0$  corresponds to the saddle point, while the energy minima are asymmetric states:
$((v,z_\tsf{A})$,  $(v,-z_\tsf{A}))$ (dotted parabolic line).
For $v\cov <v$  the energy has a single minimum at $z\ss=0$ only.

A general analysis of the explicit  $(u,p_\vp)$-representation of $\tsf{S}$ states
for the arbitrary $M$ values is given in Appendix A. In virtue of the results obtained
in Appendix A, with the aid of Eq.(\ref{Euv}) we define
the equilibrium energy of $\tsf{S}$ states as a function of the orbital momentum
\beqm{     \label{espfi}    \c E_{\,\tsf{S},u,v,M}(p_\vp) = \c E_{\ast S,u,M,{\rm sign}(p_\vp)}
\circ v\cov(p_\vp,u,M). }
By the analogy with the expression (\ref{en1}) for the energy  of $\tsf{A}$ states
we include the index $v$ which is a dummy parameter due to the condition
Eq.({\ref{indv}).
\subsection{\label{MSTAT} Minimal states in the classical limit}
In the classical limit, at fixed physical parameters ($\omega_\rho$, $\omega_z)$
of the confined system, (i.e., $v=\omega_z/\omega_\rho$ is fixed),
we search minimal energy states at a given
value of the magnetic field ($\omega_L\Rightarrow u=\omega_L/\omega_\rho$)
with respect to the integrals of motion $p_\vp,p_{\vp \ast}$.

The energy of center-of-mass motion  (\ref{cm}) is minimal at  $p_{\vp \ast}=0$.
Evidently, it does not contribute to the total
energy in the classical limit. Let $p_{\vp,\rm min}$ be a value of the orbital
momentum $p_{\vp}$ minimizing the relative motion  energy.
Eqs.(\ref{dH},\ref{uuc}) yield
\beqm{  \label{rel} 0\ss= \rho^2\, \dot \vp= p_{\vp,\rm min}- \rho_{\tsf{X}} {^{2}}\,
u,\qquad \tsf{X}=\tsf{S,A}. }
Note that for the $\tsf{S}$ states Eq.(\ref{rel}) holds only for $p_{\vp,\rm min}>0$
(see Eq.(\ref{pxM})). In virtue of this fact, with the aid of
Eqs.(\ref{rel}),(\ref{pcr}),(\ref{uvrep}), one obtains that
$v\cov=1$. Taking into account the definition of energies
Eqs.(\ref{en1}),(\ref{Euv}), we have finally
\begg{ \label{MINS}
 (p_{\vp},\rho^2, \c E_{\,\rm rel},\tsf{X})_{\rm min} = \beg{cases}{
 (0,0,\tfrac 1 2\,(1+\tfrac 2 M)\,h_{M,2\,M}(v),\,\tsf{A}) |_{M\ss=1}=(0,0,\tfrac 3 2\,v^{2/3},\tsf{A})
 &  v\le 1 \\
 (u,1, \, \tfrac 1 2\,(1 +\tfrac 2 M),\,\tsf{S})|_{M\ss=1}=(u,1,\tfrac 3 2,\tsf{S})  &  v>1,} }
Thus, in the classical limit the ground states of the confined system
(in particular, two-electron QD) exhibit diamagnetic properties in the both phases:
$\pl_B\,E_{\rm tot}\sim \pl_u\c E_{\;\rm tot}=0$.

For arbitrary values $u,v$, a collection formula (\ref{MINS}) determines
a single prescribed value of $p_\vp$ for the minimal state.
For $v<1$ the ground state is the $\tsf{A}$ minimal state. For $v\ss=1$ the ground
state  belongs to the $\tsf{A}\cap \tsf{S}$. For $1<v$ the ground state is
the $\tsf{S}$ state.

The family of minimal $\tsf{S}$ states provides a simple relation
between the strength of the magnetic field and the value of the
total angular momentum $L_z$:
\beqm{ \label{lz}   B\sim L_z \Rightarrow B=\c B_\dia\,L_z. \nonu  }
Let us define the constant $\c B_\dia$.
Taking into account that $L_z\ss=p_\vp\,L_\dia$, $B=u\,B_\dia$, $q_\dia\ss=\hbar/L_\dia$
(see, Eqs.(\ref{uvE},\ref{EBD})), and
$p_\vp\ss=u$ (see, Eq.(\ref{MINS}))   we have
\beqm{ \label{Bdia}  \c B_\dia =  B_\dia/L_\dia \quad  \Longrightarrow \quad  B_\bullet =
q_\dia\,B_\dia.     }
As a result, we obtain a magnitude of the magnetic field
$ B_ \bullet \ss=  \hbar\,\c B_\dia\ss= \Del B_{L_z \to  L_z+\hbar}$
which yields a change of the angular momentum on one Planck unit.
For $M=1$ (the Coulomb interaction) it gives
\beqm{  \label{B1class}   B_\bullet   \approx 0.724 \times
\bigl(\tfrac{m\co}{m_{\rm e}}\,\varepsilon_r\bigr )^{2/3}\,
      \bigl(\tfrac{\hbar\,\om_\rho}{\rm meV}\bigr)^{4/3} \times {\rm Tesla}.  }
\section{Vibrational corrections in the harmonic limit}
\subsection{Normal modes in the classical limit}
\label{nmcl}
As it discussed above, in physical systems, a particle undergoes small
oscillations around an equilibrium point. Let us introduce  the deviation
from equilibrium point
$\tsf{q}_\tsf{X}$ $(\tsf{X}=\tsf{A},\tsf{S})$:
$\tsf{q} \ss= \tsf{q}_\tsf{X}  + \tsf{\tilde q}$.
As a result,  the Hamiltonian function takes the form of the Taylor series
\beqm{   \label{gf}
    \c H\,(\tsf{q}_X+ \tsf{\tilde q})= \c H\,(\tsf{q}_X) +
    \SUM{\alpha}{}(\partial_{\alpha}\c H\,)(\tsf{q}_\tsf{X})\,\tilde q_\alpha+
 \tfrac{1}{2}\,\SUM{\alpha}{}\SUM{\beta}{}(\partial_{\alpha}\partial_{\beta} \c H\,) (\tsf{q}_\tsf{X})\,
  \tilde q_\alpha\,\tilde q_\beta +\ldots, \nonu  }
where the stability of the equilibrium solutions requires that the Hessian matrix
$(\partial_{\alpha}\partial_{\beta}\c H\;)(\tsf{q_0})$ should be positively defined.
Due to the axial symmetry of our system ($\dot p_\vp\ss=0$) the deviations
are considered for elements of the subset $\tsf{q}$ which form the $\rm SO(2)$
reduced phase space
$\tsf{q}\ss =\{\rho,z, p_\rho,p_z, \rho\cov,z\cov,p_{\rho\ast},p_{z\ast}\}$ (see Sec.II).
Taking into account the equilibrium solutions $\c E_{\,\tsf{X}}$
(see Eqs.(\ref{cm},\ref{en1},\ref{Euv}))
we obtain
\beqm{   \label{H12}
         \c H_\tsf{\;X} =  \c E_{\,\tsf{X}} + \c E_{\,\tsf{CM}} +
         \tfrac 1 2\,\SUM{\a}{}\,(\tilde{p}_\a{^2} +  \tl{p}_{\a\ast}{^2}) +
      \SUM{n=2}{}\c U^{\,(n)}_{\tsf{X}}(\tl \rho,\tl z) +
         \c U_{\rm CM}(\tl \rho\cov,\tl z\cov) + u\,\c E_S\,M_S+\dots ,}
where the index $(n)$ denotes the approximation order for the potential
function $\c U_\tsf{X}$.

In order to analyse the stability of the classical equilibrium,
we consider the vibrational modes in the harmonic limit $n\ss=2$
and introduce the following definitions
\beqm{
    \label{hap}
    \c U^{\,(2)}_\tsf{X}(\tl \rho,\tl z) =
      \c U_\tsf{X} =   \tfrac  1 2\,\SUM{\a,\bt}{}\,{\rm k}_{\tsf{X},\a\bt}\,\tl x_\a\,\tl x_\bt, \qquad
       \c U_\tsf{CM}(\tl \rho,\tl z) =
    \tfrac  1 2\,\SUM{\a,\bt}{}\,{\rm k}_{{\rm CM},\a\bt}\,\tl x_{\a\ast}\,\tl x_{\bt\ast}, }
For the center-of mass-motion we find
\begg{     \tsf{k}_{\rm CM}={\rm diag}\,(\Om_{\rho\ast}{^2},\Om_{z\ast}{^2}),\quad
   \label{omCM}      \binom{\Om_{\rho\ast}}{\Om_{z\ast}} = \binom{2\,\sqrt{1+u^2}}{v}.         }
For the family of $\tsf{A}$ states we have the following matrix elements
\begg{    {\rm k}_{\tsf{A}, \rho\rho}=  A  +  4\, d^{\,2}(u,v),\qquad {\rm k}_{\tsf{A},zz}=
(M+2)\, v^2- A,  \\
 {\rm k}_{\tsf{A}, \rho z} = {\rm k}_{\tsf{A}, z \rho} =
  (M+2)\,h_{M,2\,(M+4)}(v)\,\rho_A\,z_A,   }
where $ A\ss=  (2+M)\,\rho_\tsf{A}{^2}\,h_{M,2\,(4+M)}(v)$, and
$\rho_\tsf{A},z_\tsf{A}$ are given by Eq.(\ref{Aeq}).

We recall that the conditions:
Eq.(\ref{fizsol}) and $p_\vp \le p_{u,M}(v)$ (see Eq.(\ref{pcr})), -- determine
the admissible domain of $\tsf{A}$ states. At these conditions the matrix
$\tsf{k}_\tsf{A}$ is well defined
and yields the following eigenmodes $\Om_{\pm,\tsf{A}}$:
\begg{ \label{omA}  \Om_{\pm,\tsf{A},u,v,M}{^2}(p_\vp) = \tfrac 1 2\,[4(1 + u^2) + (M-2)\,v^2  \pm
         \sqrt{[4(1+u^2)-(6+M)\,v^2]^2 +\Del^2}\,], \\
       \Del^2  =  16\,(2+M)\,h_{M,2(4+M)}(v)\,d (u,v)\,|p_\vp|. }
Thus, in terms of normal modes, for the $\tsf{A}$ states we obtain
\beqm{ \label{zpm}  \c U_\tsf{A}^{\;\prime}   =
\tfrac 1 2\,(\Om_{-,A}{^2}\,\tl x_-{^2}+\Om_{+,A}{^2}\,\tl x_+{^2}),  \qquad
        \binom {\tl z_+}{\tl z_-}  =\biggl (\beg{matrix}{ \hph{-} \cos \phi  &
     \sin \phi \\ - \sin \phi & \cos \phi  }\biggr)
         \cdot \binom{\tl z_\rho}{\tl z_z},    }
where $\phi= \tfrac 1 2\,{\rm arctan}\,[2{\rm k}_{\tsf{A},\rho z}/( {\rm k}_{\tsf{A},\rho \rho}-
{\rm k}_{\tsf{A},zz})]$ and  $\tl z\ss= \tl x,\,\tl p$.

Note that the equilibrium $\tsf{S}$ states are defined by the equilibrium
parameter  $v\cov$ by means of the $v\cov$-parameterizations (see Eq.(\ref{Euv})
and the following discussion in Sec.IIIB). The expansion (\ref{hap}) for the
$\tsf{S}$ states in $(\rho,z)$-representation has a diagonal form
\begg{ %
  \label{ozor}   \tsf{k}_\tsf{S} ={\rm diag}\,(\Om_{\rho,\tsf{S}}{^2},\,\Om_{z, \tsf{S}}{^2} ), \\
  \label{omSp} \Om_{\a, \tsf{S}} \ss\equiv     \Om_{\a,\tsf{S},u,v,M}(p_\vp)=
  \Om_{\a,\ast\tsf{S},u,v,M} \circ  v\cov(p_\vp,u,M),\quad \a=\rho,z,
\intertext{where}
   \label{omS}    \Om_{z,\ast\tsf{S}}  =  \sqrt{v^2  -v\cov{^2}}, \qquad
   \Om_{\rho, \ast\tsf{S}}   = \sqrt{4\,(1+u^2)  + (M-2)\,v\cov^2}. }
Here, $\Om_{\a,\ast \tsf{S}}\ss\equiv \Om_{\a,\ast \tsf{S},u,v,M}(v\cov)$. 
Evidently, the expansion (\ref{hap}) of the $\tsf{S}$ states (which approaching
the maximal  $\tsf{A}_{u,v}$ states)
takes place around the equilibrium parameters of the confined system ($v=\omega_z/\omega_\rho)$
such as $v=v\cov$. In this case one of the normal modes  $\Om_{z,\ast\tsf{S}}\equiv0$
(see Eq.(\ref{omS})) and  it follows that
\llist{
\itt{a}  for $v\cov< v$ we have stable $\tsf{S}$ states, which we denote as $\tsf{S}_+$;
\itt{b}  the condition $v\ss=v\cov$ defines  the bifurcation point at
a given value of the magnetic field $\omega_L$, which determines the subfamily of
$\tsf{S}$ denoted as $\tsf{S}_0$;
\itt{c} the condition $v<v\cov$ defines the unstable $\tsf{S}$ states (saddle points)
denoted as $\tsf{S}_-$. }
We name the points (a,b,c) as {\sl Rules I}. In general, the condition
$v\ss=v\cov$ defines a shape (phase) transition surface in the
three-dimensional space $(u,v,p_\vp)$ (see also Fig.\ref{fig2}).
Thus, the plane $v\ss=v\cov$
divides the $\tsf{S}$ manifold on three sets $\tsf{S} \ss= \tsf{S}_-\cup \tsf{S}_0\cup \tsf{S}_+$
accordingly  to the value of parameter $\mu \ss={\rm sign}\,(v-v\cov)$.
The application of  {\sl Rules I} will be discussed in details for a particular case in Sec.V.A (see below Fig.3).

Since the maximal $\tsf{A}_{u,v}$ states define the phase transition hypersurface
$G_M(p_\vp,u,v)\ss=1$ in three--dimensional space $(u,v,p_\vp)$,
in a number of  applications it is instructive to use
$u_{\rm crit}(v,p_\vp,M)$, Eq.(\ref{ucr}). Note that for $p_\vp=0$ Eq.(\ref{ucr})
yields $u_0=u_{\rm crit}(v,p_\vp=0,M)=\sqrt{v^2- 1}$ which is fulfilled for
a standard choice of the QD parameters : $\omega_\rho \ll \omega_z$.
Thus, we conclude that for $u_0 < u$  and at the condition (\ref{pcr})
one expects the asymmetric $\tsf{A}$ states. Taking into account quantum oscillations
around the equilibrium classical ground states, we may expect a shape (phase)
transition from $\tsf{S}$- to $\tsf{A}$- states at the magnetic field strength
$u>u_0$ for  orbital momenta $p_\vp<p_{u,M}$ (see Eq.(\ref{pcr})).
Thus, if the system's parameters are subject to the condition $q_\dia<1$ ,
when the harmonic approximation is well justified, we predict a shape transtion from a lateral
to a vertical localization of two confined fermions in a magnetic field.
This general conclusion elucidates the shape transition in the excited state
found for two-electron QDs in the  magnetic field \cite{ent}. Indeed, this excited state is
formed in the local potential minimum produced by the interplay of the parabolic
three-dimensional confinement, the magnetic field and the Coulomb interaction.
\subsection{Quantization of normal modes}

To quantize normal modes of the classical Hamiltonian in the form
$h_\a (\Om_\a) =  \tfrac 1 2\,(p_\a{^2} + \Om_\a{^2}\,x_\a{^2})$
we follow the standard procedure.
The latter is  established by   means of  following  expressions
\begg{ \label{HATXP}    \binom{\hat x_a}{\hat p_a}=
\sqrt{ \frac {q_\dia} 2 }\,\binom{ \hph{-} \bar{\om}_{a} {^{-1/2}}\,
\hat B_{a+} }{-i\,\bar{\om}_{a}{^{1/2}}\, \hat B_{a-}}, \quad a=\pm,\rho,z \\
\hskip 2mm
\hat  B_{a\pm } = \pm e^{i\,\psi_a/2}\,\hat b_a{^\dag}(\bar{\om}_a) +
e^{-i\,\psi_a/2}\,\hat b_a(\bar{\om}_b). \nonumber }
The phases $\psi_\a$ provide the phase convention for states $\la \tl x_\a|n\ra$.
We fix the phases by  the conditions $\psi_\a\ss=\psi_{\a\cov}\ss=0$.

In virtue of the Poisson rules
(see, Eq.(\ref{OmN}) and the  text below Eq.(\ref{toLl}))  and
the representation (\ref{HATXP}) one obtains
\beqm{  [\hat B_{a-},\hat B_{a+}]\ss=2\,[\hat b_a, \hat b_b{^\dag} ]=2\,\del_{ab}\,. \nonu  }
Thus, the operators $\hat b_\a,\hat b^\dag_\bt$ obey the standard boson
commutation relations with respect to the boson vacuum $\hat b_\a|0\rangle=0$.
As a result we have
\beqm{   \label{hoO}  \hat h(\bar \om,\Om) =
 \tfrac {q_\dia}{4}\,(\Om^2/\bar{\om} + \bar{\om})(\hat b^\dag\,\hat b +\hat  b\,\hat b^\dag) +
        \tfrac {q_\dia}{4}\,(\Om^2/\bar{\om} -\bar{\om})( (\hat b^\dag)^2 + \hat b{^2}), \nonu   }
where $\bar{\om}$ is  a positive parameter.
The minimization  of the energy
$\c E_{\bar \om_\rho,\bar \om_z}(k_\rho,k_z)\ss= \la   \Sig_\a \hat h_\a(\bar \om_\a,\Om_\a)\ra \ss=
  \tfrac {q_\dia}  4\,\Sig_\a (\Om_a{^2}/\bar \om_\a+ \bar \om_a)\,(2\,k_\a +1)$
with respect to $\bar \om_\rho,\bar \om_z$ yields the result
$\bar \om_\a \ss=\sqrt{\Om_{a}{^2}}$. Thus, the minimum exists only for $\Om_a{^2}>0$.

It is noteworthy that the relation $L_z\ss=p_\vp\,L_\dia$ (see also Eq.(\ref{EBD}))
provides a natural quantization of the orbital momentum:
\begg{\label{qam}
L_z\ss=\frac{p_{\vp}}{q_\dia}\hbar\Rightarrow
p_{\vp} /q_\dia =    m = 0,\pm 1,\ldots,\qquad  p_{\vp\ast} /q_\dia =
m\cov= 0,\pm 1,\ldots. }
In general, the total energy has the following form
\begg{ \label{EAs}
    \c E^{\,(n)}_{\;\tsf{tot},\tsf{k},\tsf{k}\cov,M_S}(\tsf{w})=
\c E_{\,{\rm CM},u,\tsf{k}\cov } +
          \c E^{\;(n)}_{\;\tsf{X},u,v,q_\dia,M,\tsf{k}} +  u\,\c E_S\,M_S, }
\begg{
\label{XRUL}  \tsf{X} \equiv\tsf{X}(u,v,m,q_\dia,M)  = \beg{cases} { \tsf{A}  &
\text{if~}  G_M(m\,q_\dia,u,v) < 1,\\      \tsf{S}  & \text{else},} }
where $\tsf{w} \equiv (u,v,q_\dia,M)$, and the number $(n)$ defines the order
of approximation.
Here,  $\tsf{k}\ss=(k_\rho,k_z,m)$, $\tsf{k}\cov\ss=(k_{\rho\ast},k_{z\ast},m\cov)$,
where $(k_\rho,k_z,k_{\rho\ast},k_{z\ast}$)
are harmonic oscillator quantum numbers
($k_a,k_{a\ast}= 0,1,\ldots$), and $M_S\ss=0,\pm 1$ is a z-projection of the total
spin of pair electrons.

The eigenenergies of the center-of-mass motion are defined as
\begg{  \label{ECM}    \c E_{\,{\rm CM},u,\tsf{k}\cov } =
\bigl[ \sqrt{1+u^2}\,( |m\cov | +2\,k_{\rho\ast} +1) -
u\,m\cov + (\tfrac 1 2+ k_{z\ast })\,v\bigr] \times q_\dia.  }
For $q_\dia=1$ these energies  are well-known Fock-Darwin ones (see, for example, \cite{naz}).
Taking into account Eq.(\ref{qam}), we have for the energy of relative motion
in the harmonic limit ($n\ss=2$)
\begg{   \label{EX} \c E^{\;(2)} _{\,\tsf{X},u,v,q_\dia,M,\tsf{k}} =
\c E_{\,\tsf{X},u,v,M}(m\,q_\dia) +
 q_\dia \times \SUM{\a}{}\;\Om_{\a,\tsf{X},u,v,M}(m\,q_\dia)\,(k_\a+\tfrac 1 2).  }
We recall, that the classical energy $\c E_{\,\tsf{X},u,v,M}(p_\vp)$
for $\tsf{X}=\tsf{A}$, $\tsf{S}$ are defined by
 Eqs.(\ref{en1},\ref{espfi}), respectively.
The normal modes $\Om_{\a,\tsf{X}}$ for $\tsf{X}=\tsf{A},\tsf{S}$ are defined
by Eqs.(\ref{omA},\ref{omSp}),
respectively.
\section{Analysis of results}
For the sake of illustration of general results obtained for the potential
$\c{U}_M(\tsf{ r})= \alpha_M\,\tsf{ r}^{-M}$$(\alpha_{M}>0)$
we consider the most studied case of $M=1$ (the Coulomb potential).
We compare our analytical results with the numerical results obtained previously
for the three-dimensional two-electron QDs \cite{nen3} for
different $q_\dia$ values.
This analysis will allow also to illuminate the details of the
interplay between the classical and  quantum mechanical dynamics in
realistic samples.

\subsection{Classical limit}

First, let us discuss the equilibrium
classical energy $\c E_{\tsf{S},u,v,M\ss=1}(p_\vp)$ (see Eq.(\ref{espfi}))
for the $\tsf{S}$ states.
Fig.\ref{CONEV} displays the energy surface for the two-electron QD at different
values of the magnetic field $u$ and for various values of the orbital momentum $p_\vp$.
\begin{figure}
\centering{\includegraphics[scale=0.9]{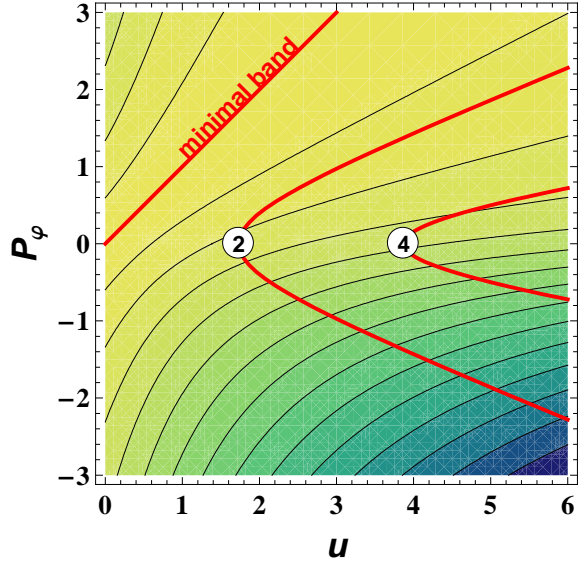} }
 \caption[$(u,p_\vp)$ projection of contours of  energy $\c E_{\,\tsf{S}}$]
 {$(u,p_\vp)$ projection of contours of  energy $\c E_{\,\tsf{S}}$.
 The energy grows from the minimal band, and each line, starting from the minimal
 band, corresponds to the increase of the energy on one unit of $E_\dia$.}
 \label{CONEV}
\end{figure}
The straight line displays the minimal energy of the $\tsf{S}$-states
in the classical limit.  According to Eqs.(\ref{rel},\ref{MINS}) this line
is obtained from the requirement
$(\pl_{p_\vp}\,\c E\;)_{\,\tsf{S},u,v,M}(p_\vp)=0$ at the condition $v\cov=1$.
The minimal states
have definite values of the orbital momentum,  which are subject to the condition
$p_\vp=u$ for $M=1$.

Once we fix the size of the QD $v=\omega_z/\omega_\rho$, the {\sl Rules I}
(see Sec.\ref{nmcl}) take place.
These rules are manifested through the hyperbolic (thick) lines $v\cov \ss= {\rm const}$
which divide  the regions of stable and unstable $\tsf{S}$ states
for various values of the orbital momentum $p_\vp$ that are available
at various values of the magnetic  field $u$. The lines are obtained by means
of the solution of the equation for different values of $u$
\begg{\label{conds} u^2 -  h_{M,4}(v\cov)\,{p_\vp}{^2}= v^2-1, \nonu  }
where  $h_{M,r}(v)$ is given  by  Eq.(\ref{HMK}).
This equation  represents  the condition: $1\ss=G_M(p_\vp,u,v)$
(see Eq.(\ref{gm}) and  Eq.(\ref{sps})) for  $h_{M,r}(v)$.

The lines are labelled by corresponding $v\cov$ numbers.
For example, for the QD size $v_1$, defined by the condition
$v\cov=1\le v_1\le v\cov=2$,
the minimal $\tsf{S}$-states are stable for
all values of the magnetic field $u$ and the orbital momenta $p_\vp$ on the
surface region $(u,p_\vp)$ restricted by the lines: "minimal band" and
$2$.
All $\tsf{S}$-states, which energies are higher those of the minimal band,
are vibrational excitations relative to the states of the minimal band.
Evidently, for $v_1$ the right region on the surface $(u,p_\vp)$
restricted by the line 2  is associated with the stable $\tsf{A}$-states, according to
the {\sl Rules I c}.

For the QD size $v_2$, defined by the condition
$v_2\le v\cov=4$, the admissible domain of values on the $(u,p_\vp)$-surface
is defined by the lower limit which is
the line "minimal band" and the upper limit which is the line $4$.
Again, the {\sl Rules I} are applied to distinguish stable $\tsf{S}$-- and
$\tsf{A}$--states for the QD size $v_2$.
\subsection{Validity of the model: a comparison with numerical calculations}
The ground state energy of a QD, as a function of magnetic
field, is studied by means of single-electron capacitance
spectroscopy  or by single-electron tunneling
spectroscopy (see for review \cite{kou}).
At low temperature $\sim 100$ mK, a large electrostatic charging energy
prevents the flow of current and, therefore, the dot has a fixed number of electrons.
Applying a gate voltage to the contacts
brings the electro-chemical potential of the contacts in resonance
with the energy $\mu(N)$ that is necessary for adding the $N$-th electron,
tunneling through the barrier, into the dot with $N-1$ electrons.
As a result, one observes experimentally kinks in the additional
energy
\beqm{  E_{\rm add}(N)=\mu (N)- \mu (N-1),  \label{aden} \nonu   }
where $\mu(N) = E(N)-E(N-1)$ is an electrochemical potential and
$E(N)$ is the total ground state energy of an $N$-electron dot.
For $N\ss=2$ we have:
$E(0)\ss=0$, $E(1) \ss=E_{\rm CM}$, $E_{\rm add}(2)\ss= E_{\rm rel}-E_{\rm CM}$.
Hence, according to Eq.(\ref{XRUL}), we have
\beqm{   \label{EAD2}  \c E^{\;(n=2)}_{\,\tsf{add},\tsf{k},\tsf{k}\cov,M_S}(\tsf{w})= -
        \c E_{\,{\rm CM},u,\tsf{k}\cov } +  \c E^{\;(n=2)}_{\;\tsf{X},u,v,q_\dia,M,\tsf{k}}
     +  u\,\c E_S\,M_S,   }
where $\tsf{X} \ss\equiv \tsf{X}(u,v,m,q_\dia,M)$ is defined in Eq.(\ref{XRUL}).
In order to define the quantum number $M_S$ we have to take into account the Pauli principle.
The spatial symmetry of wave function $\Psi (\tsf{x}_1,\tsf{x}_2)$ under the permutation of electrons
is determined  by the phase factor  $(-1)^{m+k_z}$ .
We consider a standard situation, when  the confinement in z direction is
much stronger than the lateral one, i.e., $\omega_z\gg\omega_\rho$.
Therefore, the lowest quantum numbers for the $z$-confinement are important only
for the ground state transitions of the QD in the magnetic field \cite{nen3,nen4}.
For  $k_z\ss=0$ the  total wave function is  antisymmetric, if  $(-1)^{S+m}\ss=1$. Since
$M_S=-S$,   the rule
\[ M_S\ss= -{\rm mod}_2(m),   \]
determines quantum number $M_S$ of  the minimal   antisymmetric states.

\begin{figure}[ht]
 \centering{ \includegraphics[scale=.65]{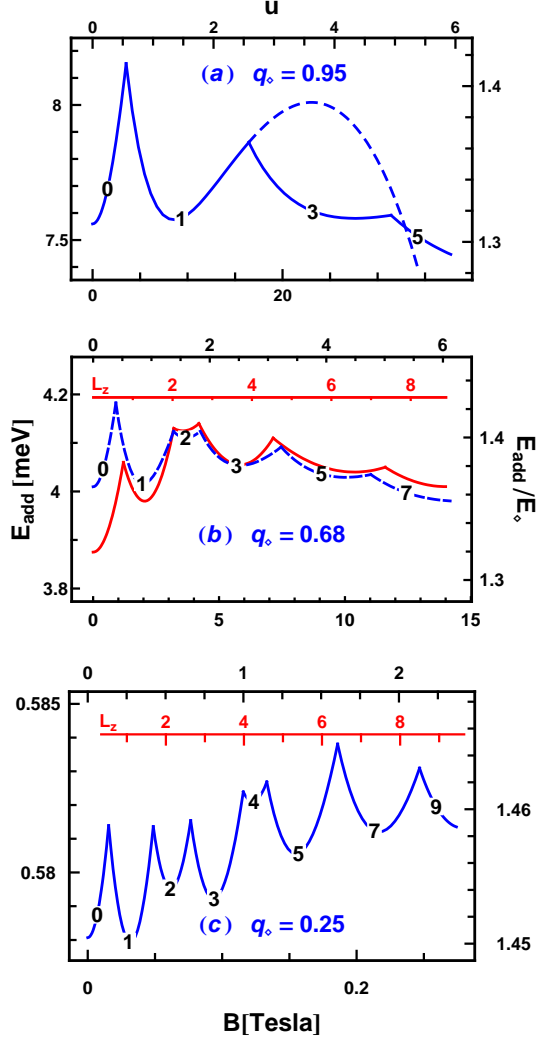}  }
\caption {
The  additional energy as a function of the $q_\dia$-factor for different values of the
magnetic field $B(u)$. The following confinement
parameters  $\hbar(\omega_\rho,\omega_z)$ are chosen: (a) 5.5,22; (b) 2,8; (c) 0.1, 4 (meV).
All calculations are done
for $m\cov =0.067\,m_e,  \eps\ss=12, |g|\ss=0.3$.
Panel (a): the evolution of $m=1$ state as function of the magnetic field, after a crossing of
the state $m=3$, is displayed by dashed line.
Panel (b): solid and dashed lines are used for the analytical and
numerical results for two-electron QD \cite{nen3},
respectively.
The straight lines in panels (b), (c) display
the values of the orbital momentum of the minimal
state as a function of the magnetic field.
 }
\label{fig4}
\end{figure}
As it is discussed in Sec.I, the  quantity $q_\dia$  defined in  Eq.(\ref{uvE}), characterizes
the strength of the quantum effects over the classical ones.
The results of calculations of the additional energies are shown
on Fig.\ref{fig4} for various values of the  $q_\dia$-factor (the Coulomb interaction).
The calculations are done at different values of the  magnetic field
$B$ (in ${\rm Tesla}$, bottom) which are related to the dimensionless
parameter $u$ (top). The left scale  counts energy in
${\rm meV}$, while  the right one expresses the energy in the $E_\dia$ unit.
The straight lines in Figs.\ref{fig4}(b),(c), demonstrate   the angular momentum scale obtained
with aid of Eqs.(\ref{lz},\ref{B1class}), and represent a classical limit.

For the Coulomb interaction, at almost equal strengths of the classical and
quantum effects (see Fig.\ref{fig4}(a)), the harmonic
approximation  fails with the increase of the magnetic field. The state with $m\ss=1$ becomes lower
than the state $m\ss=5$ at high values of the magnetic field $B>35$ Tesla.
This  behaviour of  the  energy,  defined at fixed value $v$ and  $p_\vp =p_m \ss=m\,q_\dia$,
indicates that the parameter $u$ is close to the critical value
$\bar u\ss = u_{\rm crit}(v,p_m,M)$ (see Eq.(\ref{ucr})).

To illuminate this fact we analyse the limit  $u \to \Om_{z}$ for
fixed values $v,p_\vp$.
For  $\tsf{S}$  states,  at the neighborhood of
the critical values the definition (\ref{sps}) reads :
$ 1\ss= G_M(p_m, \bar u+ z, v\cov ),\, v\cov =v + V(z)$,
where $V(0)\ss=0$   and   $ 0< \Om_{z,\tsf{S}} \Leftrightarrow z<0$.
The derivative of function $V(z)$ is obtained by means of the theorem on
implicit function:
$ 0= G_M\ord{0,1,0} dz+ G_M \ord{0,0,1} dV
\Rightarrow V\pr(0)= - G_M \ord{0,1,0} (p_m,\bar u,v)/G_M \ord{0,0,1} (p_m,\bar u,v)$.
Since the exact analytical  form of $\Om_{z,S}$ is unknown in general,
we obtain the result with the aid of the $v\cov$ parametrization
$\Om_{z,\ast \tsf{S}}$ (see Eq.(\ref{omS})),
\beqm{   \Om_{z,S}{^2} \ss =  \Om_{z,\ast \tsf{S}}{^2} \ss\approx -
       2\,v\,V\pr(0)\,\del u, \qquad\; V\pr(0)=  \frac{\bar u\,v}{v^2+z_M(p_\vp,v)}\,, \nonu  }
where  $z\ss\equiv z_M(p,v)\ss=  4\,p{^{\,2}}\,h_{M,8}(v)/(2+M) $.
If $0<\del u $ then we deal with the phase $\tsf{A}$. The expansion of
$\Om_{z,\tsf{A},\bar u+\del u,v,M}(p_m)$
for  $0<\del u$ is determined  directly from   Eq.(\ref{omA}). Comparing both
result we obtain
\beqm{ \Om_{z,\bar u +\del u,v,M}(p_m) \ss \approx  \sqrt{v\,V\pr(0)}
\times \beg{cases}{ \sqrt{ -2\,\del u } & \del u \le 0,\\
 \hfill 2\,\sqrt{\del u}\hfill  & \text{else}.} \nonu    }
Note that the  function $\del u\to -\Om_{z,\bar u +\del u, v}(p_m)$
defines curves  which are similar  to a ``$\lam$"  letter.
As a result,  the derivative $ (\pl_u  \Om_{z})_{u,v,M}(p_m)$
does not exist at $u\ss= u_{\rm crit}(v,p_m,M)$.

It proves  that the  harmonic approach  is broken down in the  domain $|\del u| \ll 1$.
More precisely, the assumption that $\c E^{\,(2)}_{\tsf{tot},\tsf{k},\tsf{k}\cov}(\tsf{w}) $  is
an  estimation of functions $\c E^{\,(\infty)}_{\tsf{tot},\tsf{k},\tsf{k}\cov,M_S}(\tsf{w}),\,\tsf{w}=
(u,v,q_\dia,M)$
presumes that  $q_\dia < \bar q_\dia(\tsf{w},\tsf{k},\tsf{k}\cov,M_S)$.
Here,  $\bar q_\dia(\ldots )$ is a radius of convergence of
series  $\Sig_{k\ss=0}^\infty\, e_k(\tsf{w},\tsf{k},\tsf{k}\cov,M_S)\,q_\dia{^k}$
obtained by means of the Taylor  expansion of the energy
$ \c E^{\,(\infty)}_{\tsf{tot},\tsf{k},\tsf{k}\cov,M_S}(\tsf{w})$.
Since coefficients $e_k(\ldots )$ have to be differentiable functions of $u$ and $v$,
we conclude that at
$\lim_{u \to  u_{\rm crit}(v,m\,q_\dia,M)} \bar q_\dia(\tsf{w},\tsf{k},\tsf{k}\cov,M_S)\ss=0$, i.e.
the considered   series do not exist at $u\ss= u_{\rm crit}(v,m\,q_\dia,M)$ for any finite value $q_\dia$.
Thus,  if the quantum numbers $\tsf{k},\tsf{k}\cov$  and parameters $v,q_\dia$
 are  fixed, the harmonic approximation does not provide a reliable description
for  $u \in [u_1,u_2]$,  where the interval $[u_1,u_2]$ contains
the point     $u_{\rm crit}(v,m\,q_\dia,M)$ and
 $\lim_{q_\dia\to 0}  u_i \ss= u_{\rm crit}(v,m\,q_\dia,M)$.
The larger  is the value of the parameter $q_\dia$
the larger is a window where the harmonic approximation
breaks down.

Evidently,  the decrease of the $q_\dia$-factor leads to the the increase of accuracy of the
harmonic approximation. We found that the analytical results describe quite well
the results of numerical diagonalization procedure \cite{nen3} with $q_\dia|_{M=1}=0.68$
(see Fig.\ref{fig4}(b)).
It appears that in these calculations  the classical effects dominate in the
dynamics of the realistic two-electron QDs under values of the  magnetic fields
available in experiments. The model allows also to trace small quantum fluctuations
in a  strong classical limit (see Fig.\ref{fig4}(c)).
In above considered cases the ground state energy is defined by $\tsf{S}$-states.

\subsection{ \label{INVARIANCE} A comparison
of additional energy spectra $E_{\rm add}$ for different $M$.}
The question remains to answer is what will happen
at the transformation $M\mapsto M\pr$ ?
The physically correct form of transformations  is obtained considering the following
symplectomorphisms  of  $(N_\dia,\Om_{N_\dia})$:
\beqm{  \hat g_{L/M}((k_1,\ldots,k_6,M),\tsf{m}_\dia)= ((k_1,\ldots,k_6,L),\tsf{m}_\dia).      }
The corresponding transformations group $ G_7 \ss=\{\hat g_a\!: a\in \Re_+\}$ obeys the following rules:
 ${\rm Ad}_{ \hat g_{L/M} }(g)\ss=g, g\ss\in  G_{N_\dia}$ (see Eq.(\ref{GNd})), $\hat g_{L/M}\in G_7$,
Moreover,  $m\co,\om_\rho,L_\dia$, hence also $q_\dia\ss=\hbar/L_\dia$,   are  $G_7$    invariant functions.

Since the map  $(N_\dia,\Om_{N_\dia})$ is not convenient to use
for the comparison of physical results, it is instructive to study the group $G_7$ as
the  transformation group of the  original  map  $(N,\Om_{N})$:
$g_a(\tsf{l},\tsf{m}_{12})\ss= \hat \kap^{-}  \circ \hat g_{a}
\circ \hat  \kap^{+} (\tsf{l},\tsf{m}_{12})$ (see
Eq.(\ref{toLl}) and the formulas above  Eq.(\ref{toLl} )).
The explicit calculation  of $g_{L/M}(\tsf{l},\tsf{m}_{12})$ gives:
\beqm{ \label{gLM}  g_{L/M} (\tsf{l},\tsf{m}_{12}) =
((l_1,\ldots,l_5,(L/M)^s\,h_{M,L-M}(z)\times l_6,L),\tsf{m}_{12}), \qquad
z = \frac {2\,l_6 }{M^s\,l_4\,l_5{^2}}.  }
We recall that   $l_6\ss=\a_M$ and $z\ss= 2\,\a_M/(M^s\,m\co\,\om_\rho{^2})$. When
the group action  is pull back   with the aid of the transformation
$(\bt,\gam) \to  (\a_M,m\co)$, determined by   Eq.(\ref{btm}), onto
the coordinates  $(\gam,\bt,M)$, one finds,
\begg{ \label{BLM1}  g_{L/M}\cdot (\gam,\bt,M)\ss=(\gam,   (L\,h_{M,-L-2})^s\,h_{M,M -L}(\gam/\bt)\,\bt,L),   }
%
In order to exhibit the group theoretical structure of this relation, we apply
the substitution  $\bt = \bt_{s,p,\gam,M}= M^s\,(\gam/\bt)^{(p-M)/(p+2)}\times \bt$,
where $-2<p$. As a result, we obtain
\beqm{  g_{L/M}\cdot (\gam, \bt_{s,p,\gam,M},M)  =  (\gam, \bt_{s,p,\gam,L},L).  }
\begin{figure}[t]
 \centering{ \includegraphics[scale=.6]{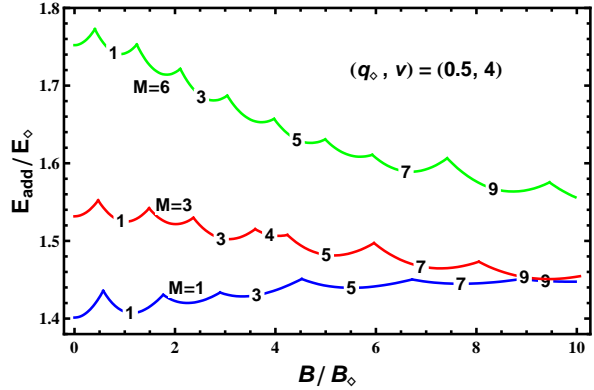} }
 \caption[  Comparison of additional energies   ]
 {The additional energies   for different potentials
 with $M\ss=1,3,6$. See text.
}
 \label{fig5}
\end{figure}
In order to compare results for different potentials,
we will study  a sequence of lists $\tsf{l}_k$ obtained by choosing a few values
$M_k$, $M_1\ss=1$ and $G_7$ action:
$\tsf{l}_{k} \ss=  g_{M_k}\cdot \tsf{l}_1$ (see Eqs.(\ref{gLM},\ref{BLM1})).
Since, the mathematical model is constructed with the aid of parameters $\tsf{k}$,
we have: $\tsf{k}_k=\hat g_M\cdot \tsf{k}_1$.

This trick enables to us to trace the evolution of  harmonic quantum  effects
with the increase of the magnetic field for different potentials on the same figure.
We consider the same parabolic confinement potential with
$v=\omega_z/\omega_\rho\ss=4$, when the quantum contribution is by twice smaller than
the classical contribution $q_\dia=0.5$. The result are displayed on Fig. \ref{fig5}.
For a better visualization, the additional energies $E_{\rm add}$ (a vertical axis)
have been multiplied by the factor $g_M\!: g_M\ss= \ss= 3/(1+M/2)$ ($g_1\ss=1$)
resulting from the formula of minimal states:
$\c E_{\,\rm rel}=\tfrac 1 2\,(1+M/2),\,1<v$ (see Eq.(\ref{MINS})).

The effect of vibrations is most visible for the potential with the larger $M$. 
Indeed, the deeper is the potential, the larger is the amplitude of vibrations.
The magnetic field diminishes quantum fluctuations. The potentials with $M=1,3$
tend asymptotically to the classical limit $\sim 3/2$ at large magnetic fields, while
the quantum fluctuations are still strong for the potential with $M=6$.
Thus, the developed model provides a  relatively simple way to analyse
a full 3D-dynamics of two fermions interacting by means of
the potential $\c{U}_M(\tsf{ r})=\alpha_M\,r^{-M}\,(\alpha_{M}>0, M>0)$
under the perpendicular magnetic field.
\section{Summary}
We formulated the algebraic approach in order to study classical and
quantum correlations in two-fermion systems confined by the 3D axially-symmetric
parabolic potential in the harmonic approximation.  The system dynamics is governed
by the interplay between the two-body interaction in the form
$\c{U}_M(\tsf{ r}) \ss= \alpha_M\,r^{-M}(\alpha_{M}>0, M>0)$, the confinement
potential and the external magnetic field. For this problem
we suggest  the  scaling symmetry $G_4$. Since the latter acts effectively
on the parameters $(B,m\cov)$, this symmetry enables one to establish
a similarity between results obtained for system Hamiltonians
with different efective masses $m\cov$. It would be desirable to test the validity
of this symmetry in experiments with real QDs.

The analytical results, obtained in the harmonic approximation,
provide a reliable description of the evolution of the ground and excited states
of two-fermion systems in the applied magnetic field.
The harmonic approximation is well justified when the classical correlations
dominate over quantum ones.
The validity of our approach have been
proved by a remarkable agreement with numerical results for
QDs parameters available in experiments \cite{nen3}.

Our analysis reveals the
coexistence of different shapes which under certain conditions may
transform from one to another.
Indeed, the interplay between classical and quantum correlations
may lead to a shape transition from a lateral to a vertical localization
of the confined electrons due to diminishing of quantum fluctuations under
certain choice of the system parameters. Such a transition is accompanied by
a spontaneous symmetry breaking of the inner parity symmetry $\c P_{\rm in}$
at the preserved integral of motion $p_\vp=${\sl const}. This general result is
nicely supported by exact numerical calculations for the case of the Coulomb
interaction \cite{ent,PR}.

\section*{Acknowledgements}
This work was partly supported by RFBR Grant No. 08-02-00118 (Russia) and
the Conselleria d' Educaci'o, Cultura i Universitats (CAIB) and FEDER (Spain).

\appendix

\section{The explicit  $(u,p_\vp)$-representation of $\tsf{S}$ states}
In order to find a general solution of Eq.(\ref{vcp}) for an arbitrary value
$M$ we recall that:
i)~the family $\tsf{A}$ states is subject to the condition (\ref{fizsol});
ii) the border line for the onset of the family maximal $\tsf{A}_{u,v}$ states is
defined by {\em Definition} \ref{TheoA}. These conditions lead to the
conclusion that  the function $v\cov$ obeys the inequality $v\cov^2< 1+u^2$
for all values $u,\,p_\vp$ identically.
To proceed further let us to introduce the following notations
\beqm{  y_u=\frac{v\cov}{\sqrt{ 1+u^2}},\qquad     P_\alpha(z)=z^{\alpha}.   }
In virtue of these definitions,   Eq.(\ref{vcp})  can written
in the following form

\begg{ \label{yuzu}    \frac{p_\vp}{b_M(u)}= P_{1/(2\,a)}\circ F_{a} \circ P_2  (y_u),
\intertext{where}
\label{defxa}   F_a(y) = y\,(1-y)^a,\quad   a =   -(M+2)/4,
\quad b_M(u)\ss= h_{M,2}\times h_{M,(M-2)/2}(1+u^2).  }
Let us precede the  analysis of Eq.(\ref{yuzu}) by the discussion of some symmetry  of $F$
identifying  $F$  with  $\tl F(a,y)\ss=(a,F_a(y))$ and assuming that  $(a,y)  \in \Re_+ \otimes [0,1]=D$.
Let $\tl f^\pm(a,y)=(a,f_\pm(a,y))$  and $\phi$ be transformation
$\phi: \tl f^\pm   \to \tl f^\mp$ given
by $\tl f^\mp(a,y)\ss= \phi( \tl f^{\pm})(a,y) \ss= (a,  f_\pm{^a} (1/a,1-y))$; hence if  $f_\pm(a,y) \in \Re_+$
then $\tl f^\pm \ss=\phi(\tl f^\mp )\ss=\phi \circ \phi(\tl f^\pm)$.
Since $\phi(\tl F)\ss=\tl F$,  so $\phi$ is the symmetry. The transformation $\phi$ decomposes:
$  \phi(\tl f)  \ss= P \circ \tl  f\circ  R$, where
 $P(a,y)\ss= (1/a,y^{1/a})$, $R(a,y)\ss=(1/a,1-y)$ and $P\circ P\ss=R\circ R={\rm id}_{D}$; hence,
inverting both sides  of equation  $\tl F_\mp  \ss=\phi(\tl F_\pm ),\,(\tl F_+\ss\equiv \tl F_-)$,
one finds
\beqm{  \tl G^\mp = \phi\cov(\tl G^\pm) =  R \circ \tl G^{\pm} \circ P  \label{GG}\,,   }
where $\tl F^\mu\circ \tl G^\mu = {\rm id}_D$. With the aid of
the original notations Eq.(\ref{GG}) transforms to the form
\beqm{\label{a5}  G^{\mp }_a(x) = (\phi\cov(G^\pm ))_a(x) =   1- G^\pm_{1/a}{^{1/a}}(x)\,. }
The geometric interpretation of components $G^\pm_a$ is obtained, taking into account
that  the derivative $F\pr$ vanishes at a point $\hat y_a\in [0,1]$; hence
$\hat x_a= F_a(\hat y_a)={\rm max}_{y\in [0,1]} F_{a}(y)$.
It proves that  the inverse  of $F$ is  determined as the doubly valued function  $(G^-_a,G^+_a)$
defined on the interval $[0,\hat x_a]$.
In order to pass to a single valued one,  let $F^\mu_a \in {\rm Map}(Y_\mu,\Re_+)$,  where
$Y_-=  [0,\hat y_a]$, $Y_+=  [\hat y_a,1]$. As a result,  $F^\pm_a$  represent two monotonic
functions, and, if $G^\mu \circ F^\mu_a= {\rm id}_{\rm Y_\mu}$,
then
\beqm{         G^-_a(0) = 0, \qquad G^+_a(0)=1, \qquad G^\pm_a(\hat x_a)= \hat y^a. \nonu  }
Note, that Eq.(\ref{vcp}) is determined as the inversion of the function
on the right hand side of Eq.(\ref{yuzu}).
It has the following form:
\begg{ \label{gv}   v_\ast(p_\vp,u,M) = \sqrt{1+u^2}\,P_{a}\circ G^-_{a}
         \circ P_{2\,a} (p_\vp/b_M(u)).
\intertext{Applying $ G^-_{a}\ss=\phi\cov(G^+)$ (see also Eq.(\ref{a5})),
one obtains}
  \label{gvbis}  v_\ast(p_\vp,u,M) = \sqrt{1+u^2}\times [ 1 -  G^+_{1/a} \circ \bar x(p_\vp)]^{1/2}, \\
    \label{xbar}     \bar x(p_\vp)\ss = [ (p_\vp/b_M(u))^{2\,a}]^{1/a} =
    (p_\vp/b_M(u))^2   < \hat x_a{^{1/(2\,a)}}.   }
where  the  parameter $a$ depends on $M$,  in  the accordance with Eq.(\ref{defxa}).
In the physical case $-2<M\Rightarrow a<0$.

Consider  the implication $\hat x_a{^{1/(2\,a)}} \ss <  z_u$ then
$x = P_{2\,a}(z_u) \in [\hat x_a,1]$ else $x\in[0,\hat x_a]$ ($a<0)$,
where $z_u\ss=p_\vp/b_M(u)$.  The equation (\ref{gv})
expresses the result  of summation  of  Taylor series for  $v\cov$
in   powers of  $1/z_u$ (i.e., for the expansion at $z_u= \infty$).
Taking  Eq.(\ref{xbar}) in the limit   $z_u{^2} -1 = \bar x(p_\vp) -1 \to 0^+$,
one observes that Eq.(\ref{gvbis}) provides the result of
summation of  the Taylor series for  $v\cov$  in   powers of
$z_u- 1$, i.e., when $b_M(u) \le   p_\vp \ll 2\,b_M(u)$.

Since Eqs.(\ref{gv}),(\ref{gvbis}), are  analytically conjugated
by means of the symmetry $\phi$, we have to calculate the function $G^-_{a}(x)$
explicitly.
Taking into account that coefficients $a_k$ of the expansion of series
$F_a(y)\ss= \Sig_{1\le k}\,a_k\,y^k$ at the point $y\ss=0$ are known,
we conclude that  the  function $G^-_{a}(x)$ can be obtained by means of
the summation of the inverse series $G^-_{a}(x)\ss= \Sig_{1\le k} \,a^\ast_k\,y^k$.
Following this way, let us write
\begg{ \label{COMP}   y \equiv  G^-_a \circ F^-_a(y)=
\Sig_{1\le k }\, a^\ast_k\,(\Sig_{1\le m } a_m\, y^m )^k=
\Sig_{1\le p}\,(\Sig_{1\le k\le p}\, a^\ast_k\,d_{[a_1,\ldots],k,p})\,y^p, }
where the coefficients $d_{[\tsf{a}],k_1,k}$ are defined by means of the relation
$(\Sig_{1\le m } a_m\,x^m)^{k_1}\ss= \Sig_{1\le p }\, d_{[\tsf{a}],k_1,k}\,x^k$.
Let $ \{ k_1,k_2,\ldots,k_r\}  \in S_k $    be a partition of number
$k\equiv\Sig_ik_i$ (an irreducible representation of the symmetric group $S_k$).
With the aid of such a partition we determine the coefficients $d_{[\tsf{c}],l,k}$ as
\beqm{\label{dcoef} d_{[\tsf{c}],l,k}  = l ! \times  \SUM{\tsf{k}\in S_k|k_1=l}{}\;
     \mbox{$\prod\limits_{i\le k}^{} $}\,c_i{^{\Del k_i}}/(\Del k_i)!, \qquad
       \Del k_i =k_i - k_{i+1}\ge 0, \;  k_{k+1}\equiv 0.           }
Here, $\Sig_{\tsf{k}\in S_k|k_1=l}$ is a sum over irreducible representations of the group
$S_k$ in which the first  member $k_1$   of $\tsf{k}$ is fixed $k_1\ss=l$. It
 constrains  the summation over the irreducible representations of the group:  $S_k\to S_{k-k_1}$.

The  summation $\Sig_{\tsf{k}\in S_k|k_1} f(\tsf{k})$ is easy to perform by
means of the operator $d_k.$  Such an operator can be defined as follow:
$\tsf{k}_{i+1}\ss= d(\tsf{k})_i,\,\tsf{k}_i\in S_k$
obtained with the aid of the following  ordering of partitions:
\begg{   \{m_1\ldots m_s  \}\ss< \{k_1 \ldots k_r  \}  \Leftrightarrow s \le r \vee (r=s) \we
    (m_s\ss< k_s \vee m_s=k_s \we  (    \ldots \vee ( m_2<k_2)\ldots)). \nonu  }
The   sequence generated by $d$ determines a list:
$\{\{k_1k\ss-k_1\},\{k_1k\ss-k_1-1\, 1\},\ldots, \{k_11^{k-k_1}\}\}$.
Evidently, the constraint $k_1\ss={\rm const}$ is consistent with
the applied ordering, which makes  this list to be complete.

Eq.(\ref{dcoef}) is valid, if $c_i\ss\neq 0$. Contrary, if  $c_{l_1}\ss=c_{l_2}=\ldots\ss=0$,
we have to put $\Del k_{l_1} =\Del k_{l_2}\ss=\ldots\ss=0 \Rightarrow c_i{^{\Del k_i}}/(\Del k_i)!\ss=1$;
hence, the elements    $i\ss=l_1,l_2,\ldots $ do not contribute to the considered product.
For $p=1$ in Eq.(\ref{COMP})  $a^\ast_1\,a_1\ss=1 \Rightarrow a^\ast_1=1/a_1$. Taking into that
$d_{[\tsf{a}],k,k} \ss=a_{1}{^k} $  (we applied:  $\Sig_{\tsf{k}\in S_k} \ss= \{k,0,\ldots,0\})$ and the relation
(\ref{dcoef}), we find  a recurrent algorithm
\beqm{ \label{RECC}  a^\ast_{i+1} \ss= \psi_i(a_1,\ldots,a_i), \qquad \psi_i(a_1,\ldots,a_i)\ss=
    - a_1{^{-i-1}} \,\Sig_{m\le i}\,a^\ast_m\,d_{[a_1,\ldots],m,i+1},   }
for the calculation of the  coefficients $\tsf{a}^\ast$. Employing this recurrent relations for
coefficients  $a_k$ of series $F_a(y)$: $a_k \ss= (-1)^{k-1}\tbinom{a}{k-1}$ one finds
\beqm{  \label{toS}  G^-_a (x) = \SUM{k=1}{\infty}\,{a^\ast_k}\,x^k\ss= x + a\,
\SUM{k=2}{\infty}\,\frac{(t)_{k-2}}{(k-1)!}\,(-x)^k,
\qquad   t =  2 - k\,(a+1), }
where $(t)_k$ is a Pochhammer symbol.

In some cases the series $G^-_a$ and $G^+_a$ studied with aid of the recurrence (\ref{RECC})
can be analytically summed, which yields the explicit  equivalence of Eqs.(\ref{gv}),(\ref{gvbis}).
Most simple form  of  $G _{M,-}(x) \ss\equiv  G^-_{a(M)}(x)$ ($G_{M,+}\ss\equiv G^+_{1/a(M)}(x))$
applied  in Eqs.(\ref{gv}) (Eq.(\ref{gvbis})) for $a\ss\equiv a(M)$ given  in Eq.(\ref{defxa})
are found in the following cases:
\begg{ G_{2,-} = (1+x)^{-1}\,x, \qquad
      G_{6,-} \ss=   (2\,x)^{-1}\,(1 + 2 x -  \sqrt{1 + 4\,x}), \nonu \\
   \label{GHG}    G_{2+\,4\,k,-}(x)  = 1 - {}_{k}
      {\rm F}_{k-1}(\tfrac1{k+1},\tfrac 2{k+1} ,\ldots,\tfrac k{k+1};\,\tfrac 2 k,\ldots,\tfrac{k-1} k,
      \tfrac{k+1}k;\,x),\qquad k=2,3,\ldots \,,}
where $ {}_{k}{\rm F}_{k-1}$ are generalized hypergeometric functions.
In particular, for $k\ss=2$ Eq.(\ref{GHG}) reduces to the form $G_{10,-}(x)=
      1-   (2/\sqrt{3\,x})\,\sinh{\tfrac 1 3\,{\rm arcsinh}\, (\sqrt{27\,x}/2)}$.
\end{document}